\documentclass[]{interact}

\usepackage[final]{changes}
\newcommand{\stkout}[1]{\ifmmode\text{\sout{\ensuremath{#1}}}\else\sout{#1}\fi}
\setdeletedmarkup{\stkout{#1}}

\usepackage{epstopdf}
\usepackage[caption=false]{subfig}

\usepackage[american]{babel}
\usepackage{csquotes}
\usepackage[style=apa]{biblatex}
\usepackage{setspace} 

\theoremstyle{plain}

\theoremstyle{definition}

\theoremstyle{remark}

\newcommand{\ind}{\mathbf{1}}

\newcommand{\iid}{\stackrel{iid}{\sim}}

\usepackage{tabularx}
\newcolumntype{E}{>{\raggedright\arraybackslash}X}
\newcolumntype{I}{>{\raggedleft\arraybackslash}X}
\usepackage{tabulary}

\usepackage{graphicx}

\author{
\name{Daryl Swartzentruber\textsuperscript{a} and Eloise Kaizar\textsuperscript{b}}
\affil{\textsuperscript{a}Centre College; \textsuperscript{b}The Ohio State University}
}

\begin{document}

\title{Small Study Regression Discontinuity Designs: Density Inclusive Study Size Metric and Performance}

\maketitle

\begin{abstract}
Regression discontinuity (RD) designs are popular quasi-experimental studies in which treatment assignment depends on whether the value of a running variable exceeds a cutoff. RD designs are increasingly popular in educational applications due to the prevalence of cutoff-based interventions. In such applications sample sizes can be relatively small or there may be sparsity around the cutoff. We propose a metric, density inclusive study size (DISS), that characterizes the size of an RD study better than overall sample size by incorporating the density of the running variable. We show the usefulness of this metric in a Monte Carlo simulation study that compares the operating characteristics of popular nonparametric RD estimation methods in small studies. We also apply the DISS metric and RD estimation methods to school accountability data from the state of Indiana.
\end{abstract}


\section*{Introduction}

The regression discontinuity (RD) design is an approach to estimating the causal effect of an intervention using observational data when treatment assignment depends on whether or not some continuous variable (known as the \textit{running variable}) falls above a cutoff. Because assignment to or eligibility for many educational interventions depends on a quantitative measure such as an assessment score, RD has gained popularity in this field since it was introduced by \textcite{Thistlewaite}. For example, \textcite{Abdulkadiroglu2014} used RD to look at the effect that attending magnet schools with entrance exam admissions had on future educational outcomes. \textcite{Lindo2010} examined the effect that GPA-based academic probation for university students had on future educational outcomes. RD designs can also be used to evaluate interventions at the school or district level. For example, many states impose sanctions on schools or districts that fail to exceed some accountability score cutoff. These educational applications often have features such as small sample sizes or sparsity around the cutoff that make RD estimation particularly challenging.

Under certain assumptions, individuals with running variable scores near the cutoff have essentially the same probability of having that score falling above the cutoff as below the cutoff \parencite{Lee&Lemieux2010}. This idea of local randomization of treatment assignment motivates two frameworks for RD estimation. The classic framework relies on a continuity assumption \parencite{hahn2001} and involves fitting separate regressions above and below the cutoff. These regressions yield estimates of the response for those with scores at the cutoff for both treatment and control groups. The difference in these fitted values is an estimate of the local average treatment effect (LATE). A more recent approach chooses a window in which local randomization is assumed to hold and conducts finite sample inference within that window. See \textcite{cattaneotitiunik22} for a review of these frameworks.

Much of the recent methodological developments in RD designs have come from the field of economics, where sample sizes tend to be relatively large. In a survey of 110 RD studies in economics journals from 1999 to 2017, \textcite{pei2021local} reports that the median sample size is 21,561. Only a third of papers have sample sizes of less than 6,000, and only three have sample sizes less than 500. Certainly some educational applications also involve large data sets. \textcite{Abdulkadiroglu2014} analyze subsets of data from around 6,000, 11,000, and 74,000 students, and \textcite{Lindo2010} have around 13,000 students in their analysis. However, many educational applications involve much smaller sample sizes, particularly if an individual educator or institution wants to estimate the effect of a local intervention. In a meta-analysis of 11 studies with 21 independent samples looking at the effect of developmental classes on future educational outcomes of college students, \textcite{valentine2017happens} found a median sample size of around 1,000, with the smallest sample having only 185 observations. Data collected at the school or district level can also result in relatively small sample sizes. For example, in this paper, we examine the effect of the threat of school sanctions from the state of Indiana on future school accountability scores \parencite{Indianainfo}. Less than 2,000 schools are evaluated on these scores, and policy makers may be interested in subsets of these data, such as the set of all high schools, that have considerably smaller sample sizes. Many popular RD estimation methods are built on asymptotic results, and simulation studies demonstrating the performance of these methods have typically used 500 as their only or smallest sample size \parencite{IK2011, CCT2014, armstrong2020simple}. The performance of these methods for smaller sample sizes has received less attention.

Related to sample size, an important consideration that has been neglected in much of the literature on RD designs is the location of the cutoff relative to the distribution of the running variable (although see \textcite{schochetpower} for an example involving design effects for clustered studies). In many educational applications, the cutoff is by design in the tail of the distribution of the running variable. The percent of students admitted to magnet schools is small \parencite{Abdulkadiroglu2014}, as is the percent of university students on probation \parencite{Lindo2010} and Indiana schools labeled as failing. The nonparametric regression techniques common for RD estimation give little or no weight to observations with running variable values far from the cutoff. Therefore, overall sample size may not be the best indicator of the RD-relevant size of a study.

This paper aims to provide practical advice for RD analysis of small data sets such as those often found in educational applications. First, we propose a simple density inclusive study size (DISS) metric that quantifies RD-relevant size better than overall sample size and is more general than other measures used in RD studies. Then we illustrate the effect of study and sample size on the operating characteristics of several leading methods of RD estimation via a simulation study. We consider methods from the continuity framework that use local polynomial regression, a nonparametric approach that relies on a tuning parameter called a bandwidth. These methods differ in their bandwidth selection algorithms as well as their inferential techniques. We consider the bandwidth algorithms in \textcite{IK2011}, \textcite{CCT2014}, and \textcite{armstrong2020simple}. We consider the robust, bias corrected inference in \textcite{CCT2014} and the notion of honest inference in \textcite{armstrong2020simple}, as well as the conventional inference referred to in both papers. In addition to these continuity methods we consider a local randomization method using Fisherian inference based on \textcite{cattaneo2015randomization}. We show that there are substantial differences in model performance among these methods for the small studies we are interested in. Since our DISS metric does not depend on a particular RD estimation method, researchers are able to use it to help choose a method that may work well for their specific study size, and it is also useful to methodological researchers developing and evaluating new approaches to small sample RD estimation.

The rest of the paper is organized as follows. We provide an overview of the RD estimation methods we consider. Then we introduce our proposed DISS metric, followed by a presentation of the results of our simulation study. We then apply the study size metric and the considered RD estimation methods to the Indiana school accountability data before closing with a discussion and our recommendations.

\section*{Popular RD Estimation Methods} \label{methods}

We define $X$ to be the running variable, with a cutoff at $X=c$. We focus here on the ``sharp" RD case, where the treatment $D$ is assigned solely based on whether or not the value of the running variable is at or above the cutoff, i.e., $D=\ind_{[X\geq c]}$. Let $\mu(x)=\mu^-(x)\ind_{[X<c]} + \mu^+(x)\ind_{[X \geq c]}=E(Y|X)$ be the true underlying mean function, with $\mu^-(x)$ and $\mu^+(x)$ the mean function below and above the cutoff, respectively. The LATE of interest is $\tau=\mu^+(c)-\mu^-(c)$. In the continuity framework this treatment effect can be estimated by plugging in fitted values from separate regressions on either side of the cutoff to obtain $\hat{\tau}=\hat{\mu}^+(c)-\hat{\mu}^-(c)$. 

\textcite{Thistlewaite} used global linear regression for their original RD estimation. To provide more flexibility, some researchers have used higher order global polynomials instead. However, in these higher order models, points far from the cutoff have large influence relative to points close to the cutoff \parencite{Gelman2019}, which is an undesirable property given the desire to minimize the bias of the estimated mean function at the cutoff. Thus local nonparametric techniques, and in particular the local polynomial regression technique used by the continuity methods considered in this paper, may be superior. The estimated mean function in local polynomial regression is a weighted combination of the response values, using a weighting function called a kernel that relies on a bandwidth as a tuning parameter. See \textcite{fan1996local} for more details. The choice of kernel function from amongst those commonly employed, such as the triangular and Epanechnikov functions, typically makes little difference in the estimated mean function because they produce similar weights. Different bandwidth values, on the other hand, can lead to larger differences in weights and therefore larger differences in the estimated mean functions. Many data-driven bandwidth selection algorithms have been proposed in the literature, and these algorithms represent one of the key differences among the continuity methods we consider. Since these methods fit separate regressions above and below the cutoff, some algorithms produce a separate bandwidth for each regression. The algorithms considered for this paper, however, select a single bandwidth for both regressions. Another key difference among the continuity methods in this paper is the inferential techniques employed by each. We address these two important features in the following sections and then describe an alternative approach based on local randomization.

\subsection*{Bandwidth Selection Algorithms} \label{bw selection}

One popular bandwidth selection approach for local polynomial regression is to choose a bandwidth that optimizes an estimated objective function of interest.
\textcite{IK2011}, henceforth IK, use as their objective function an asymptotic approximation of the mean squared error (AMSE) of their RD estimator. Minimization of their AMSE expression with respect to the bandwidth, $h$, leads to an infeasible optimal bandwidth $h_{IK}$. They estimate the unknown quantities in this expression to yield a fully data-driven bandwidth 

\begin{equation} \label{hoptik.hat}
    \hat{h}_{IK}=\left[\frac{1}{n}\left(C_K\right)\left(\frac{\hat{\sigma}^2_+(c)+\hat{\sigma}^2_-(c)}{\hat{f}(c)(\hat{\mu}_+''(c)-\hat{\mu}_-''(c))^2+\hat{r}(c)}\right)\right]^{1/5}
\end{equation}
where $C_K$ is a function of the kernel, $\hat{\mu}_+''(c)$ and $\hat{\mu}_-''(c)$ are the estimated second-order one-sided derivatives of the underlying mean function above and below the cutoff, respectively, $\hat{\sigma}^2_+(c)$ and $\hat{\sigma}^2_-(c)$ are the right and left hand limits, respectively, of the estimated error function $\hat{\sigma}^2(x)=\hat{Var}(Y|X)$ at the cutoff, and $\hat{f}(x)$ is the estimated density of the running variable. IK also include a regularization term, $\hat{r}(c)$, that is not present in the infeasible expression, to help prevent similar estimated second derivatives from leading to extremely large bandwidth values.

\textcite{CCT2014}, henceforth CCT, use the same AMSE and infeasible optimal bandwidth from IK as their starting point, although they generalize the expressions to allow for different model and inferential choices and use a different regularized algorithm to calculate their data-driven bandwidth $\hat{h}_{CCT}$.

\textcite{armstrong2020simple}, henceforth AK, perform RD inference based on the assumption that the mean functions on either side of the cutoff lie in a class with second derivatives globally bounded by $M$. The AK approach effectively replaces $(\hat{\mu}_+''(c)-\hat{\mu}_-''(c))^2$ in $h_{IK}$ with $4M^2$ to obtain an infeasible optimal bandwidth
\begin{equation} \label{hoptak}
    h_{AK}=\left[\frac{1}{n}\left(C_K\right)\left(\frac{\sigma^2_+(c)+\sigma^2_-(c)}{4f(c)M^2}\right)\right]^{1/5}.
\end{equation}
Their data-driven bandwidth $\hat{h}_{AK}$ does not incorporate regularization and estimates the bandwidth by directly minimizing the finite sample mean squared error (MSE).  AK contend that the choice of $M$ should be made \textit{a priori} to maintain the honesty of their confidence intervals. However, they also consider a data-driven estimate of $M$, denoted $\hat{M}$, which may be valid in some cases.  

\subsection*{Inferential Procedures} \label{inference}

Currently popular RD methods involve using the chosen bandwidth to fit local polynomial regressions to the left of the cutoff to estimate $\mu^-(x)$ and separately to the right of the cutoff to estimate $\mu^+(x)$. Recall that the difference in these estimated mean functions at the cutoff is the estimated LATE, $\hat{\tau}$. The conventional (CV) inferential procedure pairs this point estimate with an estimate of the standard error of $\hat{\tau}$ to calculate a confidence interval:
\begin{equation}
    I_{CV}=\hat{\tau}_{CV}\pm z_{\alpha/2}SE_{CV}(\hat{\tau}).
\label{CV_I}
\end{equation}
CCT and AK both use a nearest neighbors approach to calculate conventional standard errors, $SE_{CV}(\hat{\tau})$. 

CCT mention two concerns with the conventional approach: non-negligible bias and empirical coverage well below the nominal level. They propose a robust bias-corrected (RBC) inferential procedure designed to address these concerns. They estimate the bias using higher order local polynomial regression and subtract that estimate, $\hat{B}$,  from the conventional point estimate. They also use a correction term $C$ in their variance estimate to account for the extra variation that comes from bias estimation. This yields the interval
\begin{equation}
    I_{RBC}=\left(\hat{\tau}_{CV}-\hat{B}\right)\pm z_{\alpha/2}\sqrt{SE_{CV}^2(\hat{\tau})+C}=\hat{\tau}_{RBC}\pm z_{\alpha/2}SE_{RBC}(\hat{\tau}).
\end{equation}

AK present an alternative approach to improving coverage based on the idea of honest confidence intervals. They simply inflate the critical value in the conventional interval (Equation \ref{CV_I}) by using a critical value $z^*_{1-\alpha}(t)$ from a folded normal reference distribution with scale parameter 1 and shape parameter $t$ based on an estimated worst case scenario. This procedure produces what AK call a fixed length confidence interval (FLCI):
\begin{equation} \label{flci}
    I_{FLCI}=\hat{\tau}_{CV}\pm z^*_{1-\alpha}(t)SE_{CV}(\hat{\tau}).
\end{equation}
Despite the increased critical value, they show their intervals actually tend to be shorter than those produced by $I_{RBC}$ while maintaining proper coverage.

\subsection*{Local Randomization Methods} \label{locrand}
The continuity methods above use asymptotic expressions in their bandwidth algorithms and/or inferential techniques, which raise concerns for small samples. An alternative approach put forth by \textcite{cattaneo2015randomization}, henceforth CFT, conducts finite-sample inference in a window close to the cutoff. The key assumptions of their local randomization approach are that treatment status in this window is as good as randomly assigned and that the response values in the window depend on the running variable only through treatment assignment (i.e., the underlying mean functions in the window are flat). 

CFT follow typical methods of randomization inference, such as using the Fisherian sharp null hypothesis of no treatment effect. Under this hypothesis, within the chosen window the response values are fixed quantities that do not depend on the treatment assignment. Thus the treatment effect for each individual is zero. Note that this hypothesis implies the implicit null hypothesis in the continuity approach, which says the average treatment effect $\tau$ is zero but says nothing about the individual treatment effects. 

With an appropriately chosen randomization mechanism, the distribution of a test statistic, such as difference-in-means, under the sharp null hypothesis is known. Thus the p-value for the test can be calculated as the sum of the probabilities of all possible sets of treatment assignments whose test statistic is at least as extreme as the one that was observed. This p-value can also be approximated based on a random selection of possible treatment assignments. 

Further assumptions are required in order to obtain point and interval estimates of the treatment effect. Specifically, CFT propose the use of a local constant treatment effect model, in which the treatment effect is the same for each observation in the cutoff. Then a confidence interval for this treatment effect can be calculated by finding the values of the hypothesized treatment effect in which the test is not rejected. 

The choice of window in which to perform such inference is the main challenge of applying local randomization methods to RD designs. CFT propose a window selection based on the idea that treatment assignment should be unrelated to the response inside the window. Their method involves a set of nested, symmetric windows in which the balance of the covariates is tested, starting with the largest window and shrinking it until a window is found in which balance cannot be rejected. They also propose setting a minimum number of observations in each window. They suggest a minimum of ten based on power considerations.

\section*{Study Size Metric} \label{m.sec}

As with most statistical techniques, we expect the quality of bandwidth estimation and RD inference to degrade as sample size shrinks. For very large studies, the differences in performance among the various bandwidth algorithms and inferential techniques discussed may be somewhat small, especially between those that are based on the same asymptotic formulas. For small studies, however, we would expect larger differences in performance. We would like to provide advice to practitioners on which methods work well for small studies, as well as develop a framework researchers can use to compare methods in simulation studies. In order to do so we need a size metric.

Our goals make certain features in a size metric desirable. Since we are interested in RD methods that estimate local average treatment effects, the metric should quantify the size relative to the cutoff. In order to compare competing RD methods in a simulation study or recommend a method for a particular application, the metric should not be based on a specific method. Furthermore, as the goal is to characterize a study instead of an analysis, the metric should allow for multiple response variables, including those that may not have been collected yet.  

These features disqualify certain commonly used size metrics. Clearly the overall sample size does not account for the local nature of our estimation techniques. A small data set that has most of its observations around the cutoff may lead to better inference than a much larger data set that has most of its observations far from the cutoff. An alternative size metric that is often reported in RD studies is the number of observations within a single bandwidth of the cutoff. This is a reasonable approach, as kernels popular for RD local polynomial regression, such as the triangular and Epanechnikov kernels, give zero weight to observations more than one bandwidth value away from the cutoff. However, the bandwidth used to calculate this metric is typically the one that is used for the RD analysis, meaning a choice in RD estimation method has been made and a particular response variable has been selected.

We propose a density inclusive study size (DISS) metric similar to the alternative metric above except that it uses a bandwidth that does not rely on a choice of estimation technique or response variable. Specifically, we use the classic rule of thumb bandwidth from \textcite{silverman1986density},

\begin{equation} \label{silverman pop}
    h_{ROT}(s^*)=0.9(s^*)n^{-1/5}.
\end{equation}
Here $s*$ is the minimum of the sample interquartile range (IQR) of the running variable divided by 1.34 and the sample standard deviation of the running variable. Thus our sample-level DISS metric, $m$, is the number of observations within this bandwidth of the cutoff,
\begin{equation}
    m=\sum_{i=1}^n \ind_{[c-h_{ROT}(s^*)\leq X_i \leq c+h_{ROT}(s^*)]}.
\end{equation}

Towards the goal of a framework for comparing RD methods using simulation studies, it is also helpful to quantify the size of a study at the population or generating model level. We do so with $\bar{m}(n)$, the expected number of observations within one bandwidth of the cutoff for a given overall sample size $n$:
\begin{equation}
    \bar{m}(n)=nP(c-h_{ROT}(\sigma^*)<X<c+h_{ROT}(\sigma^*)).
\end{equation}
This uses a population level version of the Silverman bandwidth:
\begin{equation} 
    h_{ROT}(\sigma^*)=0.9(\sigma^*)n^{-1/5}.
\end{equation}
Here $\sigma^*$ is the minimum of the population IQR of the running variable divided by 1.34 and the population standard deviation of the running variable. In the next section we utilize $\bar{m}$ as the size metric in our simulation study. 

\section*{Simulation and Results}
We use an extensive Monte Carlo simulation study to explore the small-sample performance of common RD estimation methods and demonstrate the usefulness of our proposed DISS metric. Our simulation settings vary across four dimensions: estimation method, running variable distribution, study size, and underlying mean function.

For our continuity methods we consider the three data-driven bandwidth selection algorithms from the earlier section, based on IK, AK, and CCT. For each bandwidth algorithm, point estimates and confidence intervals are constructed based on three different inferential techniques: CV, RBC, and FLCI. We use local linear regression for both the bandwidth algorithms and inferential techniques. We consider all combinations of bandwidth algorithm and inferential technique, for a total of 9 continuity methods. 

For the AK bandwidth algorithms and the FLCI technique we use the data-driven estimate $\hat{M}$ as we expect this may be more common in practice. In Section SA2 of the Supplemental Appendix we present results from using the true value of $M$ calculated from the underlying mean function as a sensitivity analysis. All continuity methods use the triangular kernel. 

We also consider the local randomization (LR) methods described earlier, with a sharp null hypothesis and difference-in-means test statistic. The data we are using in our empirical application does not have any covariates with which to implement CFT's window selection algorithm. Therefore we choose a window based on the minimum number of observations idea from CFT. In unreported simulations we found that their suggested minimum of 10 observations was generally outperformed by a minimum of 5 for the study sizes considered. Therefore in these results we present only the latter, which we refer to as LR5.

We perform all calculations in R (v\replaced{4.4.0}{4.1.1}; \cite{R}). We implement the IK and CCT algorithms and the CV and RBC techniques using the R package \textit{rdrobust} (v\replaced{2.2}{2.1.1}; \cite{rdrobust}). We implement the AK algorithm and FLCI technique using the R package \textit{RDHonest} (v\replaced{1.0.0}{0.4.1}; \cite{RDHonest} \added{and v0.3.2; \cite{RDHonestold}}). We implement the CFT methods using the R package \textit{rdlocrand} (v1.0\replaced{;}{,} \cite{rdlocrand}).

\begin{figure}[t!]
    \centering
    \includegraphics[width=\textwidth]{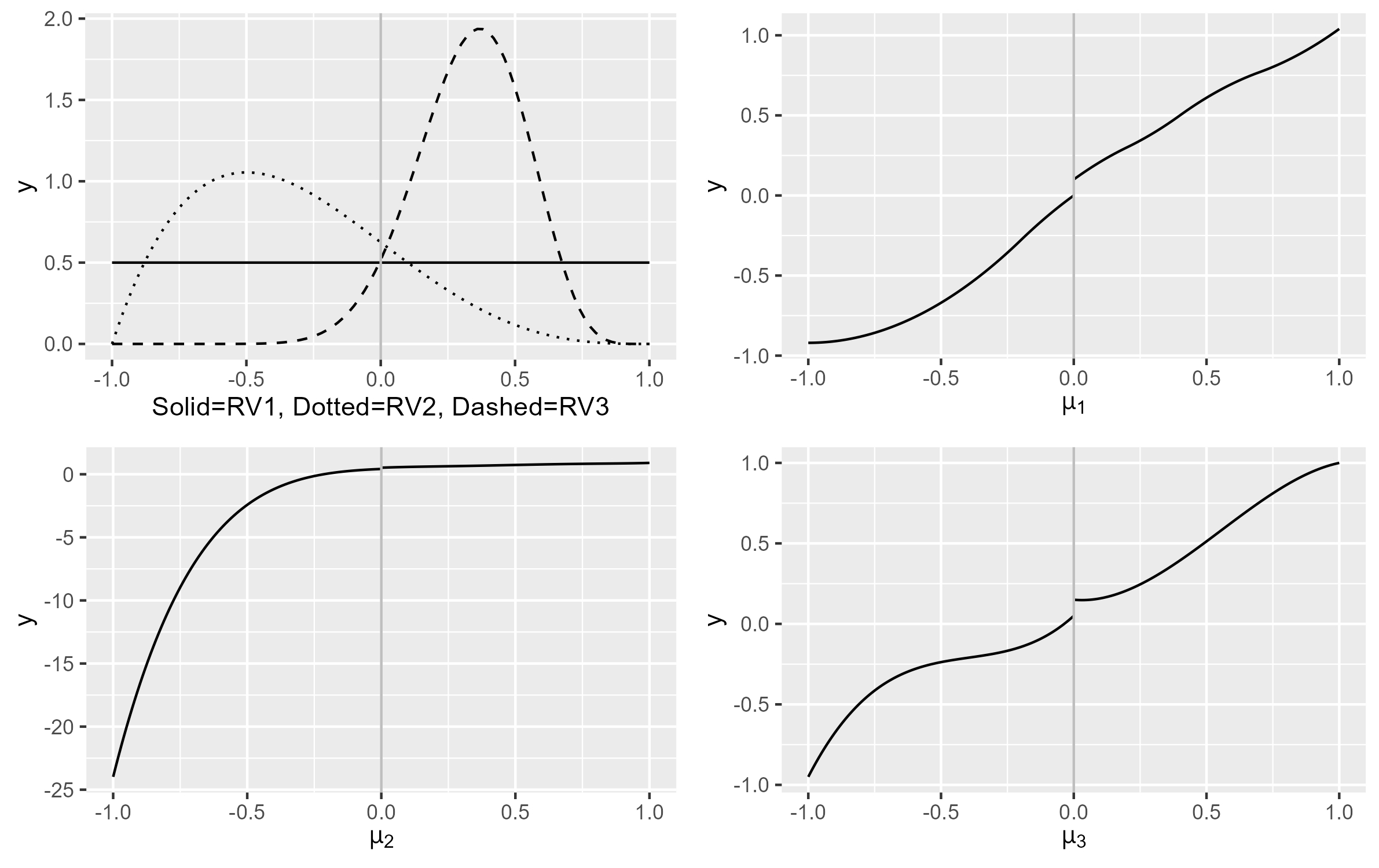}
    \caption{Running variable densities and underlying mean functions. The vertical gray line represents the cutoff.}
    \label{dgp1}
\end{figure}

To generate values $X_i$ of the running variable, we follow the simulation studies in IK and CCT in first independently drawing values $Z_i$ from a Beta distribution and then applying a transformation $X_i=2Z_i-1$. Thus the support of the transformed distribution of the running variable is $[-1,1]$. The cutoff is set to be $c=0$. IK and CCT use the same Beta distribution for all design settings. However, we are interested in examining the effects of different levels of sparsity around the cutoff, and choose three different Beta distributions to aid in this analysis. The first running variable, denoted RV1, has a transformed Beta(1,1) distribution, and has half of its density on either side of the cutoff. This underlying distribution was used in the simulation study of AK. The second running variable, RV2, has the transformed Beta(2,4) distribution from IK and has approximately 19$\%$ of its density above the cutoff. The third running variable, RV3, has a transformed Beta(14,7) distribution and has less than 6$\%$ of its density below the cutoff. This distribution is modeled after the Indiana school accountability data discussed in a later section. The upper left panel of Figure \ref{dgp1} depicts these densities. 

To determine the sample sizes for our simulated data sets we start by choosing reasonable values of our population DISS $\bar{m}$. We then calculate the sample size needed to achieve those values for each of the running variable distributions, which we report in Table \ref{mn}. Note that we have two sample sizes, $n=140$ and $n=354$, that appear once in the table for each running variable distribution. This allows us to make comparisons across values of $\bar{m}$ for a fixed value of $n$ in addition to comparisons across $n$ for a fixed value of $\bar{m}$.

We consider three different underlying mean functions $\mu(x)$. The first, referred to as $\mu_1$, is a modified version of the mean function from Design 3 in the simulation of AK. That simulation explored inference at a point and not regression discontinuity, so a jump at the cutoff has been added to the original mean function, which consists of quadratic splines with knots away from the cutoff. Thus the second derivatives immediately on either side of the cutoff are equal. The second derivative bound of this function is $M=\replaced{2}{6}$. The second mean function, $\mu_2$, comes from Design 3 in the simulation of IK and is based on a modification of the data from \textcite{lee2008randomized}. It consists of quintic polynomials on either side of the cutoff that differ only in their intercept. This function has more curvature than $\mu_1$, particularly below the cutoff, with a second derivative bound of $M=233.26$. The third mean function, $\mu_3$, is based on the Indiana data. It consists of separate cubic polynomials on either side of the cutoff, with differing second derivatives and a bound of $M=16.2$. All three mean functions have a vertical discontinuity of 0.1 at the cutoff and are continuous elsewhere. The equations of the three mean functions are 
\begin{align}
        \mu_1(x)& =(x+1)^2-2s(x+0.2)+2s(x-0.2)-2s(x-0.4)+\added{2}s(x-0.7)-0.92 \nonumber \\
        & \quad +(0.1)\ind_{[x\geq 0]} \\
        \mu_2(x)&=0.42+0.84x-3.0x^2+7.99x^3-9.01x^4+3.56x^5 + (0.1)\ind_{[x\geq 0]} \\
        \mu_3(x)&=(0.05+1.5x+3.2x^2+2.7x^3)\ind_{[x<0]} + (0.15-0.15x+2.5x^2-1.5x^3)\ind_{[x\geq0]} 
\end{align}
where $s(x)=(x)_+^2=$max$\{x,0\}^2$ is the square of the plus function. Graphs of these functions are included in Figure \ref{dgp1}.

We consider all pairs of a running variable density and a mean function, which gives us nine data generating processes (DGP). We denote these by concatenating the designations for the running variable and mean function, e.g., RV1$\mu_1$. The response values $Y_i$ are generated from the values of the running variable $X_i$ as $Y_i=\mu(X_i)+\epsilon_i$, where $\epsilon_i \iid N(0, .1295^2)$. This error distribution was also used in the IK simulation. We generate 50,000 simulated data sets for each DGP at each value of $\bar{m}$. 

We analyze the performance of the methods in four categories. For small sample sizes, bandwidth algorithms and inferential techniques do not always produce finite estimates, so an important consideration in evaluating methods of RD estimation is simply how often those methods work. Thus we look at bandwidth success rates, the percentage of repetitions in which an algorithm produces a finite bandwidth value. In this section we also examine the distributions of the calculated bandwidth values, as the size of the bandwidth plays an important role in the subsequent operating characteristics. Second, we look at the interval estimate success rate, the percentage of repetitions where a method produces a finite confidence interval. Third, we evaluate the performance of finite point estimates using the bias, empirical standard error (EmpSE), and mean squared error (MSE). Fourth, we evaluate interval estimates using the median interval width and empirical coverage. Throughout, we use the RV2$\mu_2$ DGP as a baseline to describe typical patterns across methods, then highlight deviations across the various DGPs. 

\subsection*{Bandwidth Calculation} \label{success}

One concern for RD estimation with small sample sizes is the ability to produce finite estimates. To produce a finite estimate a method first must produce a finite bandwidth. Thus we begin our analysis by looking at the success rate of the included bandwidth algorithms and the distributions of the bandwidths they produce. At study sizes of $\bar{m}=21$ and above, all bandwidth algorithms had success rates at or near 100$\%$. However, for the smallest study size of $\bar{m}=10$, \added{some of} the algorithms begin to struggle. The CCT bandwidth algorithm struggles the most, producing finite bandwidth values for fewer than 83$\%$ of iterations for RV2$\mu_2$\replaced{, while the IK and AK algorithms produce finite values more than 98\% of the time}{ compared to more than 99$\%$ of iterations for IK and AK}. This trend continues for other DGPs as seen in Table SA1 in the Supplemental Appendix. 

\begin{figure}[t!]
    \centering
    \includegraphics[width=\textwidth]{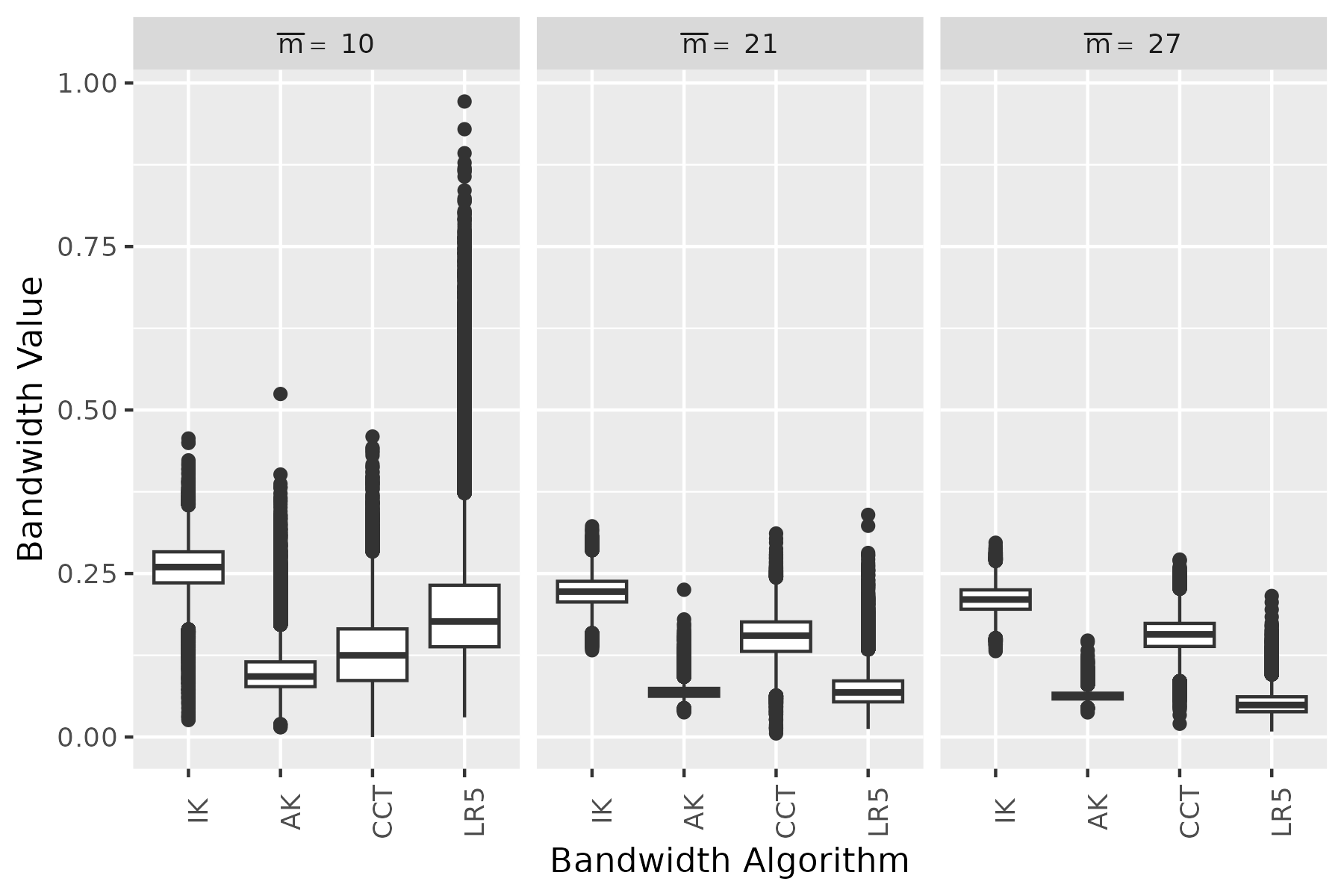}
    \caption{Distribution of bandwidth and window values for RV2$\mu_2$.}
    \label{plot.bw1}
\end{figure}

The bandwidth values themselves are important in understanding the the performance of the estimation methods as we will see. Figure \ref{plot.bw1} includes the distributions of all finite bandwidth values produced by the given algorithms\added{ for RV2$\mu_2$}, as well as the distribution of window values for the local randomization method. (This means that the distributions have different numbers of observations due to differing success rates.) We see that regardless of study size $\bar{m}$, the IK algorithm tends to produce the largest median bandwidth values followed by the CCT and then the AK algorithms. Recall that the IK and CCT algorithms are designed to optimize the same AMSE, but for these small studies Figure \ref{plot.bw1} demonstrates that the different approaches lead to substantially different bandwidth values. 

We are also interested in the relationship between study size and bandwidth/window distribution. As study size increases, the bandwidth distributions for each algorithm naturally become less variable and have fewer outliers. Given the $n$ in the denominator of Equations \ref{hoptik.hat} and \ref{hoptak} and the direct relationship between $\bar{m}$ and $n$, we also expected an inverse relationship between study size and median bandwidth value. This inverse relationship is present for IK and AK, although the median CCT bandwidth value remains roughly constant across study sizes. The median LR5 window value also shrinks as study size increases, as with more data there are more values closer to the cutoff. For the smallest study size the LR5 value is typically wider than both the AK and CCT bandwidth values, but as study size increases the median LR5 value drops below the median bandwidth values for all algorithms. The patterns relating bandwidth/window values and study size for the other DGPs are similar to the ones here for RV2$\mu_2$, as seen in Supplemental Appendix Figures SA1-SA4.

\subsection*{Interval Estimate Success Rates} \label{interval success}

Even for those iterations where a finite bandwidth is calculated, there is no guarantee that RD estimation methods will produce finite point or interval estimates. The continuity methods rely on fitting separate local polynomial regressions on either side of the cutoff, which can be impossible if there is not enough data ``close" to the cutoff, as quantified by the bandwidth value. (Calculating the bias estimates for the RBC technique and the standard errors for all methods can be even more challenging.) Thus, methods using algorithms that tend to produce larger bandwidths, such as IK, tend to have higher success rates. Table \ref{tab1} gives these rates for the RV2$\mu_2$ DGP, defined as the percentage of simulated data sets that produce finite bandwidth/window values and interval estimates. Methods using the IK algorithm \added{tend to} have the highest success rates among the continuity methods for all study sizes, with rates of at least \replaced{97}{90}$\%$ even when $\bar{m}=10$. \added{The only method that produces a higher success rate is AK/FLCI.} As study size goes up the \added{continuity} methods achieve near universal success, indicating that with enough data even methods using relatively small bandwidth values can still produce finite interval estimates. \added{The local randomization methods, conversely, are able to achieve near universal success even for small study sizes.} These patterns hold for all DGPs as seen in Tables SA2-SA4 in the Supplemental Appendix.

\deleted{Methods employing FLCI inference have much lower success rates than those using the other inferential techniques. This may indicate that it is difficult to estimate the worst case scenario needed for the inflated critical value in Equation \ref{flci}. This seems to be particularly true for relatively small bandwidth values, as the AK/FLCI method has a success rate of around 12$\%$ for $\bar{m}=10$ and less than 80$\%$ for $\bar{m}=20$. Conversely, one advantage of the LR methods is its high success rate, as it is able to perform inference even for very small study sizes. }

These success rates also demonstrate the value of our DISS metric relative to the overall sample size $n$. For example, the DGPs RV1/$\bar{m}=27$ and RV3/$\bar{m}=10$ \replaced{both}{all} have the same overall sample size of $n=140$. However, while RV1/$\bar{m}=27$ has near universal success for both $\mu_1$ and $\mu_3$, \replaced{more than}{nearly} half of the methods have success rates lower than \replaced{two-thirds}{50$\%$} for $\mu_1$ and $\mu_3$ with RV3/$\bar{m}=10$, as seen in Tables SA2-SA4 in the Supplemental Appendix. The RV3/$\bar{m}=21$ DGP has an overall sample size of n=351 and yet still has lower success rates than the RV1/$\bar{m}=27$ DGP for most methods despite that much larger sample size. For a fixed mean function, success rates in our simulation tend to be more similar when they have the same value of $\bar{m}$ than when they have the same value of $n$.

\subsection*{Point Estimation} \label{point}

Comparing methods in terms of point estimation (and later interval estimation) is made more challenging by the differing success rates. To provide a fair comparison, we summarize only iterations that produce finite interval estimates for all methods for a particular DGP and value of $\bar{m}$. This restriction is most problematic for $\bar{m}=10$, where \replaced{in some cases less than half of the iterations are considered}{the vast majority of iterations are not considered (recall the low success rates for AK/FLCI)}, but for larger study sizes we are less concerned about any differences in analysis due to this restriction. Note that the CV and FLCI point estimates are theoretically the same, and while there are minute differences that exist between the results due to algorithmic implementations, we choose to present only the CV results here. 

\begin{figure}[t!]
    \centering
    \includegraphics[width=\textwidth]{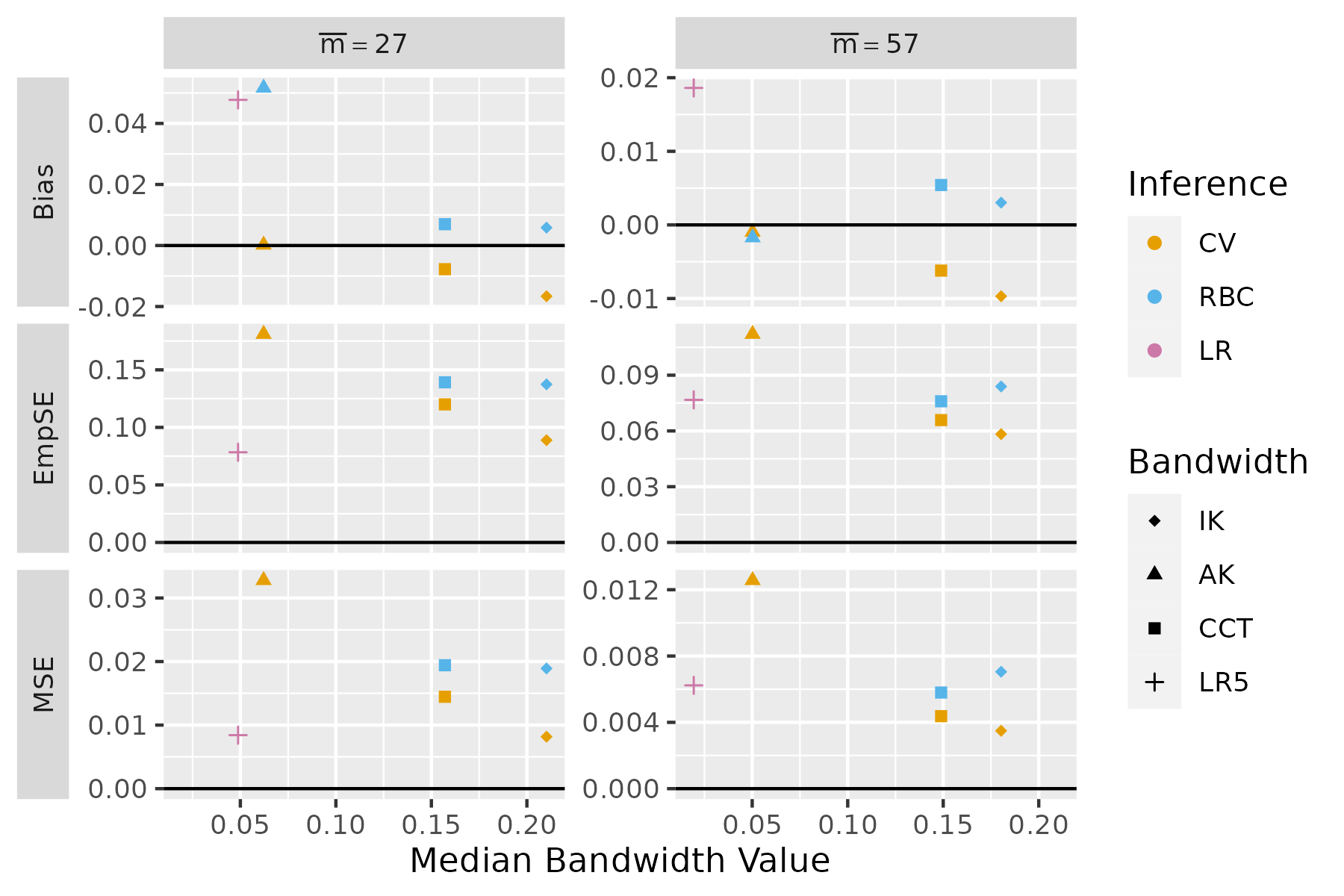}
    \caption{Performance measures for methods using CV, RBC, and LR inference for RV2$\mu_2$ at two study sizes. \textit{FLCI values are essentially the same as CV and thus omitted. Values based on iterations in which all estimates where finite: 99.3$\%$ of all iterations for $\bar{m}=27$ and greater than 99.99$\%$ for $\bar{m}=57$. The graph omits EmpSE and MSE values for AK/RBC for $\bar{m}=27$ (14.3 and 203.6, respectively) and $\bar{m}=57$ (0.20 and 0.04). The maximum Monte Carlo standard errors (MCSE) for the values shown are 0.0008 (Bias) and 0.0006 (EmpSE and MSE).}}
    \label{point.scatter1}
\end{figure}

Bias is certainly a concern for small sample RD estimation. Most of the methods have an absolute relative bias of more than 5$\%$ for RV2$\mu_2$ when $\bar{m}=27$, although there is less bias when $\bar{m}=57$. These bias values are given in Figure \ref{point.scatter1} along with EmpSE and MSE values. RBC methods are effective at reducing the bias for some, but not all, of the bandwidth algorithms. The cost of the attempted bias correction is larger empirical standard errors for all algorithms. This allows the CV(FLCI) methods to have lower MSE values than the RBC methods for all algorithms in this DGP. In particular, RBC methods that use the AK algorithm, which tends to produce small bandwidth values, have very high EmpSE and thus very high MSE. This may indicate that for small study sizes, the extra estimation needed to do bias correction may not be worth it, especially when paired with relatively small bandwidths. Among the CV(FLCI) methods, those paired with IK have the lowest MSE, with CCT also being somewhat competitive. However the small bandwidth values produced by AK result in relatively poor point estimation performance, even for CV(FLCI) inference.

The bias of the LR estimates is higher than those of the continuity methods for this DGP. However, the LR method makes up for this bias with low EmpSE, including the lowest of all methods for $\bar{m}=27$. These low EmpSE allow the LR method to be competitive in terms of MSE, particularly for the smaller study size. The above findings are also supported by Figure SA5 in the Supplemental Appendix, which gives the results for the other study sizes in our simulation.

\begin{figure}[t!]
    \centering
    \includegraphics[width=\textwidth]{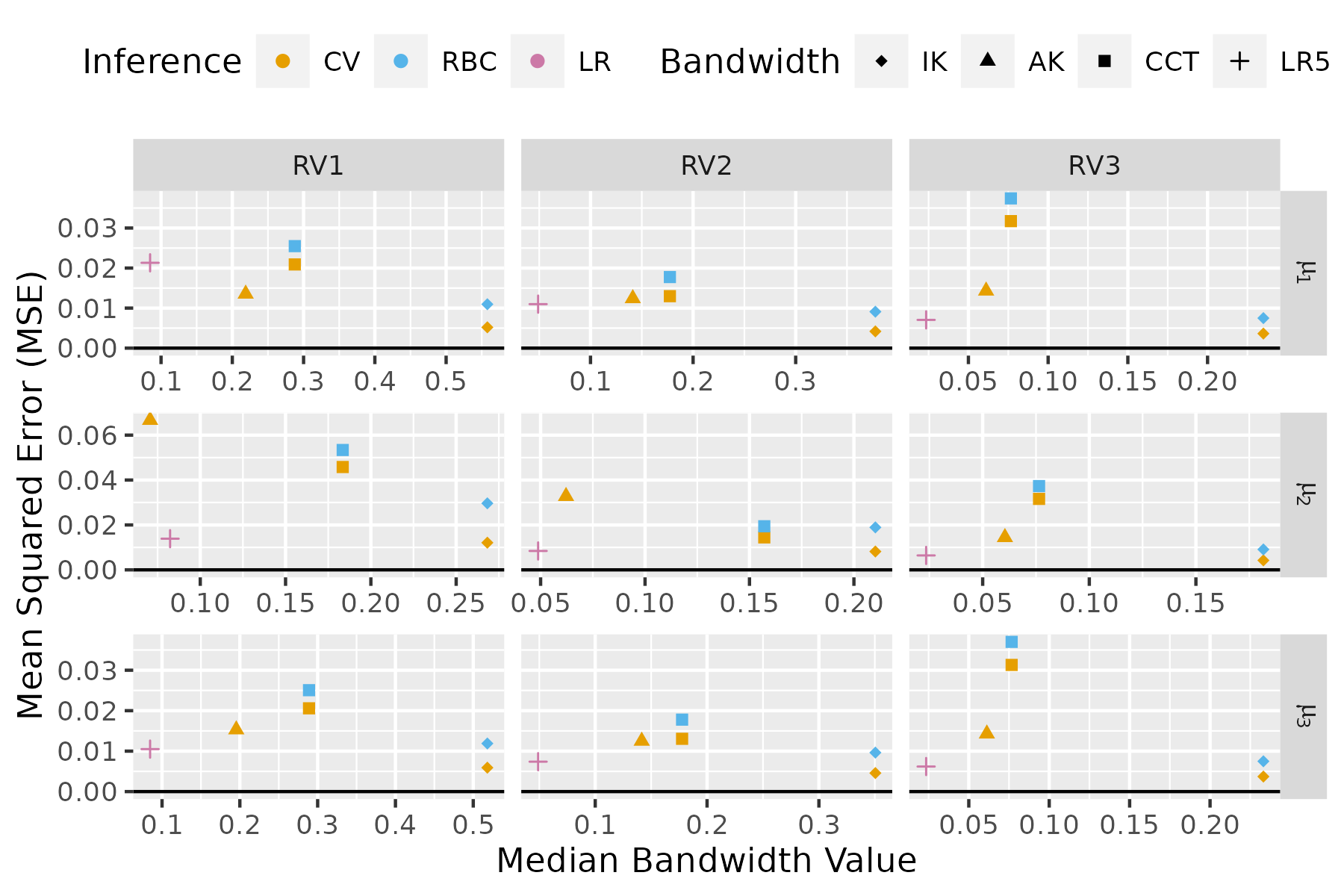}
    \caption{MSE of point estimates for different methods and designs at $\bar{m}=27$. \textit{FLCI values are essentially the same as CV and thus omitted. Values based on iterations in which all estimates were finite, which was at least 99$\%$ of iterations except for RV1$\mu_2$ (93$\%$). Values of AK/RBC are omitted, which range from 0.04 to 0.21 except for RV1$\mu_2$ (409.7) and RV2$\mu_2$ (203.6). The maximum MCSE are 0.01 for RV1 and RV3, and 0.0006 for RV2.}}
    \label{point.scatter.mse}
\end{figure}

The tendency of methods using CV(FLCI) to dominate those using RBC in terms of MSE extends to the other considered DGPs, as seen in Figure \ref{point.scatter.mse} for $\bar{m}=27$, and is again due to large EmpSE values for methods using RBC not being mitigated by enough bias reduction (see Supplemental Appendix Figures SA6-SA7). The IK/CV(FLCI) method has the lowest MSE among the continuity methods for all DGPs at this study size. One notable difference among DGPs is the relative performance of the AK/CV(FLCI) method. For RV2$\mu_2$ this method performed quite poorly, but it is much more competitive in terms of MSE for several of the other DGPs, particularly in situations where the median AK bandwidth value is relatively closer to those from the other algorithms, such as RV3. 

The performance of the LR method also depends on the DGP. Recall that the LR method is built on the assumption that the underlying mean functions are flat inside the window. While this typically implausible assumption is not met in any of our mean functions, we might expect the LR method to perform worse for more serious violations of this assumption. Among the functions considered here, $\mu_1$ is perhaps the least flat around the cutoff, being relatively steep on both sides. For $\mu_2$ and $\mu_3$ the LR method has the second lowest MSE behind IK/CV(FLCI), but for $\mu_1$ it does not perform as well, possibly due to the relative flatness of the three functions. Note that we do see an interaction between mean function and running variable distribution. LR still performs fairly well for RV3$\mu_1$ despite the steepness, but the RV3 density curve below the cutoff is strongly skewed to the right, which limits the expected spread of the observed response variable within the window. For small studies then, it seems that the performance of LR methods may suffer in some scenarios when the local randomization assumption is not at least approximately met.

With our design settings we can again examine the usefulness of the DISS metric by comparing DGPs with the same overall sample size but different running variable distributions, resulting in different values of $\bar{m}$. RV1/$\bar{m}=27$, RV2/$\bar{m}=21$, and RV1/$\bar{m}=10$ all have a sample size of $n=140$. The MSE tends to increase as the value of $\bar{m}$ decreases in these DGPs for the continuity methods, as seen in Figure SA8 in the Supplemental Appendix. The difference is most pronounced between the RV2 and RV\replaced{3}{1} DGPs, in which the former has an $\bar{m}$ value more than double that of the latter. Figure SA9 in the Supplemental Appendix shows a similar pattern for three DGPs that have an overall sample size of $n=354$. Note however that the MSE of the local randomization methods stays relatively the same for all values of $\bar{m}$ in these comparisons. This is expected, as the minimum observation criteria of our window selection method leads to relatively similar window values and thus relatively similar MSE values. We might see more of a relationship between $\bar{m}$ and MSE for LR inference with a different window selection method like the CFT approach mentioned in the section on local randomization.

\subsection*{Interval Estimation} \label{interval}
To compare interval estimation we look at the empirical coverage for the nominally 95$\%$ confidence intervals, as well as the median interval width. By construction the CV intervals will have the smallest median widths among continuity methods for a given bandwidth value, discounting minor differences in algorithmic implementation, since both the RBC and FLCI techniques widen the intervals. The FLCI technique uses an inflated critical value but is centered around the same point estimate as the CV interval, resulting in higher empirical coverage. The RBC technique adds a term to the standard error to account for the variability in the bias estimation, but since it is centered in a different place it may have higher or lower empirical coverage than the other techniques. LR inference is fundamentally different from the continuity techniques, and thus has no theoretical ordering in width or coverage relative to the others. The top performing methods should provide reasonable coverage with the smallest possible interval widths. 

\begin{figure}[t!]
    \centering
    \includegraphics[width=\textwidth]{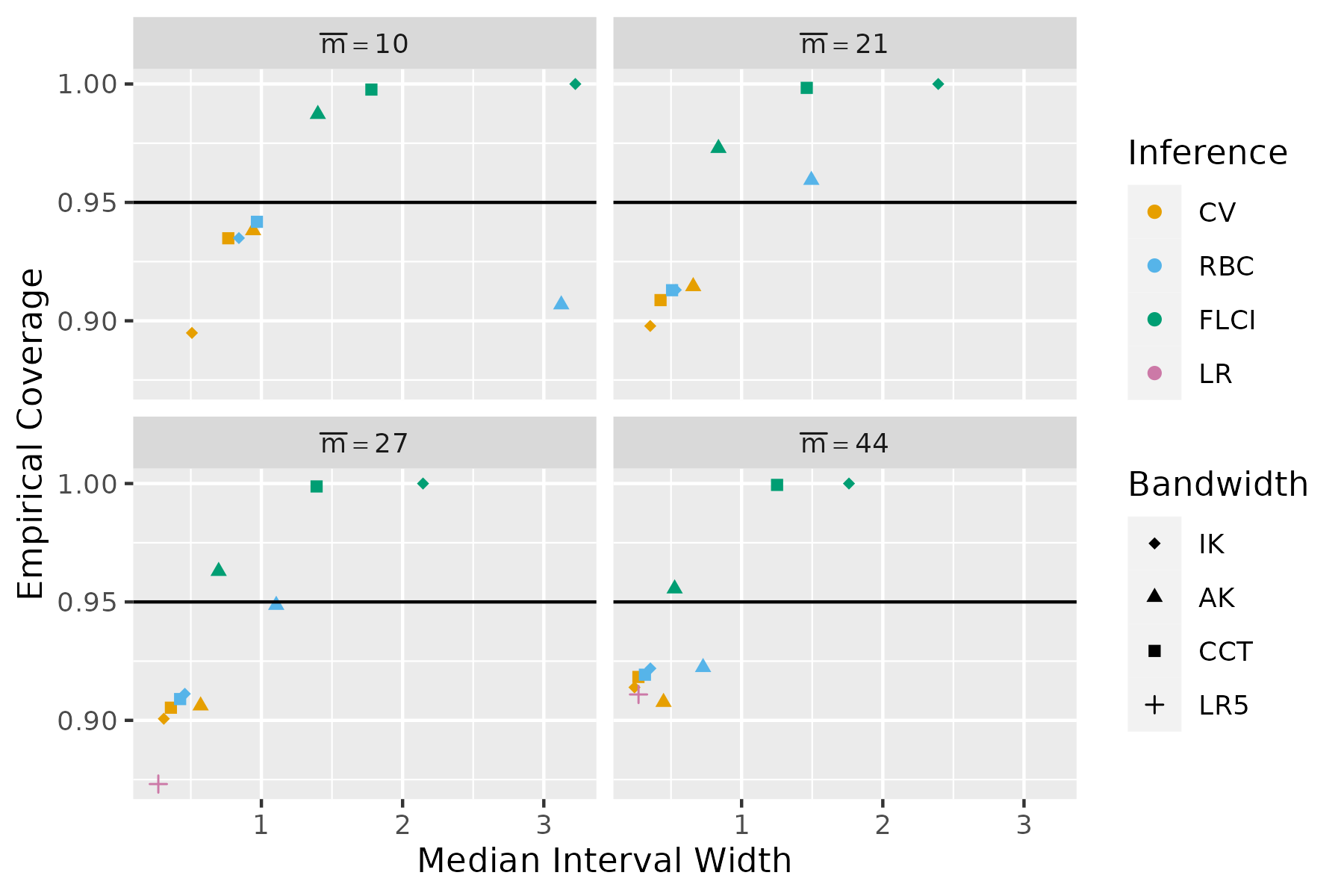}
    \caption{Empirical coverage and median interval width for all methods and four study sizes with RV2$\mu_2$. \textit{Values based on iterations in which all estimates where finite: 46.9$\%$, 96.6$\%$, 99.3$\%$, and 99.99$\%$ for $\bar{m}$ values of 10, 21, 27, and 44, respectively. The graph omits values for LR/LR5 for $\bar{m}=10$ (median interval width of 0.33, coverage of 0.53) and  $\bar{m}=21$ (0.27, 0.82). The maximum Monte Carlo standard errors for the coverage estimates shown are 0.002 ($\bar{m}=10$) and 0.001 ($\bar{m}=21$, $\bar{m}=27$, and $\bar{m}=44$).}}
    \label{coverage1}
\end{figure}

For RV2$\mu_2$ at all study sizes, the intervals typically produced by the methods using FLCI are so wide that these methods may be undesirable despite their high coverage, as seen in Figure \ref{coverage1}. However, the AK/FLCI method yields the smallest median interval width from among the FLCI methods and still has coverage above the nominal rate, making it a reasonable overall choice. The CV methods do not achieve the nominal coverage rate, but their smaller median widths ensure that they are competitive overall. In particular the IK/CV method consistently yields quite narrow intervals and yet has reasonable coverage at larger study sizes. The RBC methods do improve coverage compared to the CV methods for a given bandwidth algorithm, but these gains may not be substantial enough to warrant the added interval width. Finally, the LR methods produce the narrowest intervals for the three smallest study sizes considered, but have extremely poor coverage compared to the other methods for those study sizes. For the larger study sizes the LR method has more reasonable coverage but no longer produces the narrowest intervals.

\begin{figure}[t!]
    \centering
    \includegraphics[width=\textwidth]{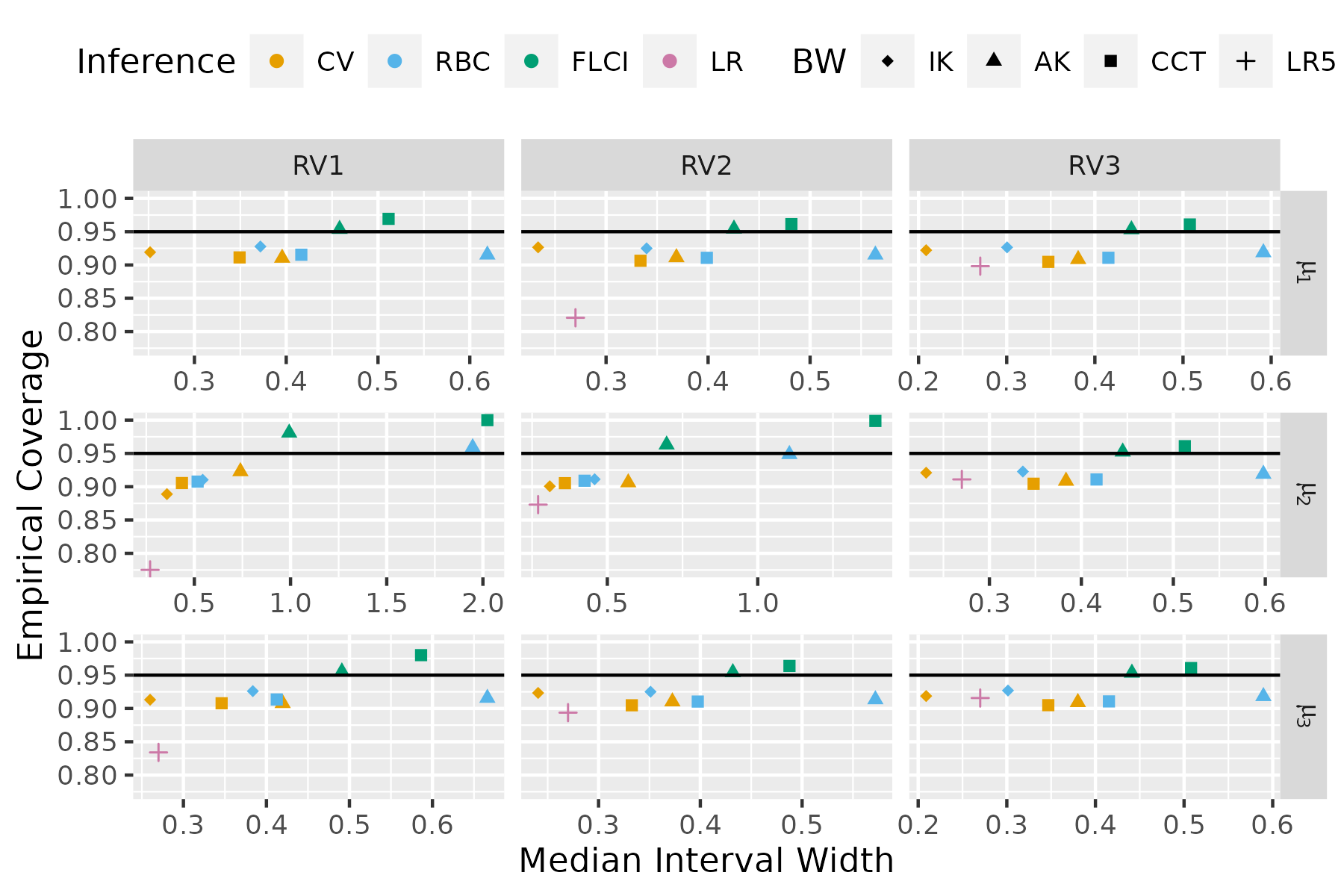}
    \caption{Empirical coverage and median interval width at $\bar{m}=27$ for all DGPs. \textit{Values based on iterations in which all estimates were finite, which was at least 99$\%$ of iterations except for RV1$\mu_2$ (93$\%$). The graph omits values of LR5/LR for RV1$\mu_1$ (median width of 0.30, coverage of 0.65) as well as IK/FLCI, which have coverages greater than 0.998 and widths at least 50$\%$ higher than the next highest value. The maximum Monte Carlo standard errors for the given coverage estimates are 0.002.}}
    \label{coverage2}
\end{figure}

Many of the patterns we see for RV2$\mu_2$ continue for the other DGPs, as seen for $\bar{m}=27$ in Figure \ref{coverage2}. FLCI intervals substantially improve coverage compared to CV intervals, but at the cost of very wide intervals, particularly in DGPs where the estimated curvature is high, such as those with $\mu_2$\deleted{ and those with RV3}. RBC intervals provide only marginal gains (and occasionally losses) in coverage that are typically not enough to justify the extra width. The IK/CV method continues to yield narrow intervals and competitive coverage, and along with AK/FLCI is one of the top performing methods in terms of the width-coverage tradeoff. The LR method also produces consistently narrow intervals. However, as we saw in the previous section, the LR point estimates are quite biased for several of the DGPs. This leads to very poor empirical coverage, lower than 85$\%$ for nearly half of the DGPs. In a majority of the DGPs the LR method is dominated by that of IK/CV, with the latter having higher coverage and smaller median interval widths. 

Figure SA10 in the Supplemental Appendix looks at the relationship between coverage and interval width for the same DGPs for the larger study size of $\bar{m}=57$, and we see similar patterns to those that exist for the smaller study sizes. The primary difference is that the LR methods tend to have relatively better coverage but much wider intervals for several of the DGPs at the larger study size, and are only competitive in a couple of scenarios. \added{However if we increased the size of the window we may see improved performance at this study size.}

As with point estimation, we can compare properties of interval estimation for DGPs with the same overall sample size but differing values of $\bar{m}$. In Figures SA11 and SA12 in the Supplemental Appendix, we see that in these scenarios as $\bar{m}$ decreases the median interval widths tend to increase. However the coverage values often stay relatively the same or even decrease with this decrease in $\bar{m}$. Thus when considering the width-coverage tradeoff, having more observations near the cutoff tends to lead to better interval estimation for a fixed sample size, just as it does for point estimation.

In sum, we found that the IK/CV methods consistently produced the narrowest intervals while typically only sacrificing a small amount (typically less than 5$\%$) of coverage even for the smallest DISS values. As such, we recommend its use in most cases for small to moderate studies where $m$ is less than around 60.  In such studies where exact coverage is required, we recommend the typically wider AK/FLCI intervals. As the sample size (and thus DISS) grows across our simulated datasets, the practical distinctions between the results from various methods begin to shrink.  As such, for studies with moderate to large DISS values, we may prefer the AK/FLCI interval that strictly controls the confidence level for studies.

\section*{Real Data Example}\label{indiana}

In 2015 the state of Indiana revamped their K-12 school accountability system \parencite{Indianainfo}. The new system, which began with the year 2015-2016, gives each school an accountability score between 0 and 120 based on a various performance and growth measures. Those scores are then translated into A-F letter grade categories. A school that scores below a 60 is automatically given an F. The state imposes various consequences if a school receives an F for a certain number of years. These consequences vary based on the type of school and the number of consecutive years the school has received an F. 

This situation can be thought of as a sharp regression discontinuity design. We choose the running variable to be the 2017 school accountability scores \parencite{Indianadata2017} and the response variable to be the 2018 school accountability scores \parencite{Indianadata2018}. These are the first two years after the new system was implemented for which full data are available. Note that the state does not implement sanctions based on just one year of earning a failing grade, so the treatment can be thought of as the threat of sanctions or the lack of such a threat that occurs when a school is given a failing grade, rather than the sanctions themselves. Schools with 2017 scores greater than the cutoff do not receive the threat of sanctions, so to be consistent with the definition of $\hat{\tau}$ given earlier the LATE estimated in this section will be the effect of not receiving the threat of sanctions. Thus a negative value of this LATE would indicate that the threat of sanctions is improving scores, which is presumably what the state of Indiana would like to be the case. 

Indiana schools are categorized according to type and grade level for the purpose of the accountability system. School types include \textit{traditional public}, \textit{charter}, and \textit{choice} schools. The latter two categories will be grouped together here as \textit{private} schools for the sake of convenience, even though there is differing opinion as to how charter schools should be categorized. In terms of grade level, \textit{low} schools contain at least some of the K-8 grades, while \textit{high} schools contain at least some of the 9-12 grades. Note that a K-12 school is considered part of both categories, so there is some overlap. Subsets of the overall set of Indiana schools can be formed based on combinations of school type and grade level. These subsets may be of interest as it is possible that the effect of the threat of sanctions may be different across subsets. These subsets also highlight differences in sample sizes and sparsity around the cutoff that is relevant to our discussion of RD LATE estimates for small samples. Table \ref{subset} contains the overall sample sizes, sample sizes below the cutoff, and values of $h_{ROT}$ and $m$ for the overall data set and selected subsets. 

In 2017 only 88 schools in the data set were labeled as failing, representing less than five percent of the total of 1933 schools. This percentage is even smaller for some of the subsets of interest. This data set provides a good example of an education application with a cutoff in the tail of the running variable distribution, and thus care must be taken when considering the overall sample size. In the group of all Indiana schools, only 51 are within a Silverman bandwidth of the cutoff, and yet the sample size of 1933 is much larger than those DGPs in Table \ref{mn} with $\bar{m}=$57. The subset of high schools is larger than that of private schools, but the latter has more values below the cutoff and more values within a rule of thumb bandwidth of the cutoff, providing a useful test case to compare our DISS to overall sample size. 

We use the methods from our simulation study to obtain interval estimates for the RD treatment effect for each of the above subsets. We choose again to use the data driven choice of $\hat{M}$ for our second derivative bound for the AK bandwidth algorithms and FLCI inference. We are using data from a new accountability system and thus do not have the luxury of historical data that might give us a different value of $M$ to use instead. For the LR window we use the same 5 observation minimum to be consistent with the simulation settings. 

\begin{figure}[t!]
    \centering
    \includegraphics[width=\textwidth]{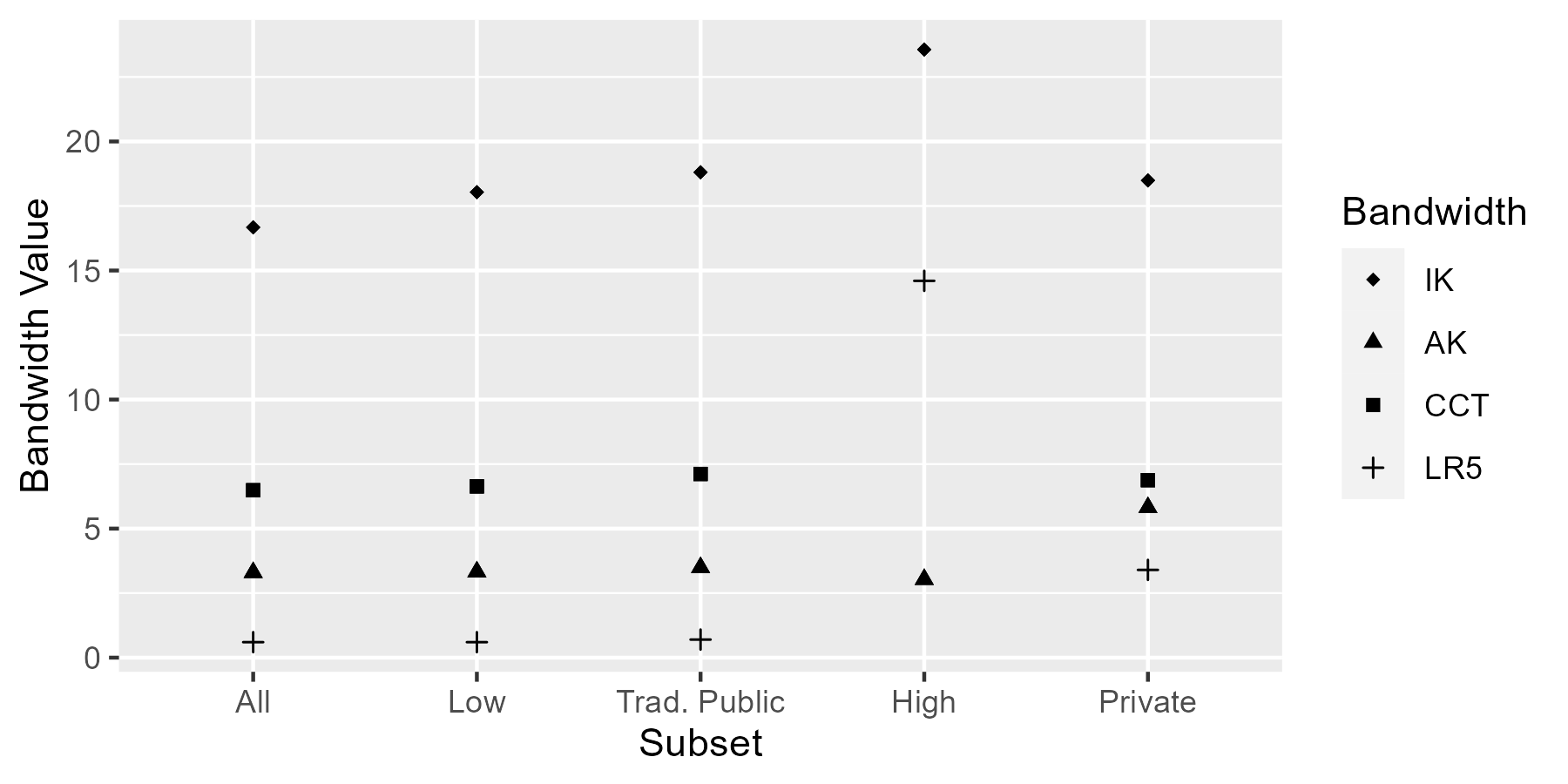}
    \caption{Bandwidth values for different Indiana subsets.}
    \label{subset.bandwidths}
\end{figure}

The IK bandwidth values are the highest for each of the considered subsets by a substantial margin, followed by CCT and then AK, as seen in Figure \ref{subset.bandwidths}. The LR5 window is smaller than all of the bandwidths for the larger subsets, but becomes much larger for the \replaced{subset of high schools, which has the smallest value of $m$}{smaller subsets}. This pattern is consistent with that in the simulation study. Note that the CCT algorithm is unable to produce a bandwidth for the set of high schools, but all algorithms produce bandwidths for the set of private schools. There are 101 more high schools than private schools, but the set of private schools has a larger value of $m$. These extra observations near the cutoff likely play a large role in the success of the bandwidth algorithms.

\begin{figure}[t]
    \centering
    \includegraphics[width=\textwidth]{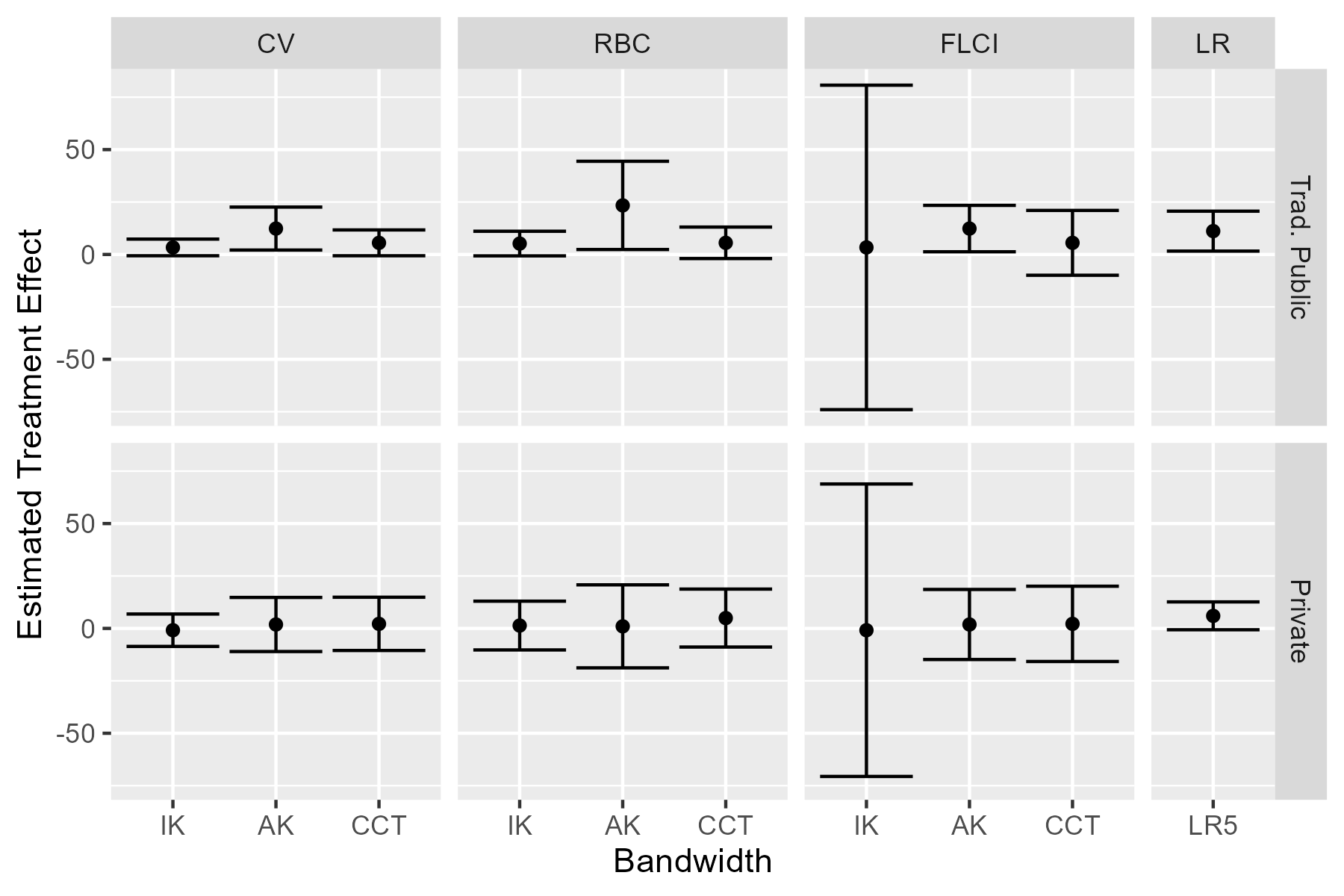}
    \caption{Treatment effect estimates and 90$\%$ confidence intervals for select Indiana subsets.}
    \label{indiana.ci}
\end{figure}

Figure \ref{indiana.ci} gives estimated treatment effects and 90$\%$ confidence intervals for two mutually exclusive subsets of Indiana schools. The set of private schools represent a low study size of $m=10$, while the traditional public schools have a more moderate study size of $m=42$. For the set of \replaced{traditional public}{private} schools, all of the treatment effect point estimates are positive and four of the ten are statistically significant. Those latter methods show some evidence that being labelled a failing traditional public school impacts future performance measures, but in the opposite direction than the state would likely intend. For the set of private schools, there are no statistically significant results, and \replaced{one}{two} of the point estimates \replaced{is}{are} negative. These results may indicate differences in the effect of the treatment among different subgroups, but as no comparative methods were used we must be careful not to overstate this case.

The relationships between method and interval width in these results mimic those of the simulation study in several key ways. Recall that the two top continuity methods from the simulation study were IK/CV and AK/FLCI. For both the traditional public and private schools, AK/FLCI produces a wider interval with likely exact coverage, but suggests that larger effect sizes are consistent with the data. Conversely, IK/CV rules those larger effect sizes out, though the method may be ruling too many large effect sizes out in a way that drops the coverage a bit below 90$\%$.

Another important pattern reinforced by these results is the potential bias of the local randomization methods for small study sizes. For the set of private schools, the LR5/LR method has the largest point estimate and nearly excludes 0 from its interval, whereas all of the other intervals have lower bounds well below 0. It is plausible, then, that the LR5/LR point estimate is so biased that its interval is the only one that fails to include the true value of the treatment effect. For the larger set of traditional public schools, however, the LR5/LR point estimate is between that of IK/CV and AK/FLCI and the LR5/LR interval would not appear to be in the same danger of failing to cover the true effect.

Figure SA13 in the Supplemental Appendix shows results for the other Indiana subsets of interest with similar patterns to those mentioned above. Additionally, we see that several methods\deleted{, including AK/FLCI,} are unable to obtain a finite interval estimate for the set of high schools, while all methods obtain a finite interval estimate for the set of private schools with a smaller overall sample size but larger $m$. Also, there is little difference between the effect estimates for the set of low schools and the set of all schools. The latter includes more than 250 schools than the former, but all but two of them are above the cutoff and most are considerably above the cutoff, leading to the same value of $m$ for both sets. Thus adding large numbers of observations away from the cutoff may change the overall sample size but may not have a noticeable effect on the estimation. These results highlight the usefulness of the DISS metric in characterizing the size of an RD study.

Ultimately we failed to see a statistically significant negative effect across the five subsets of Indiana schools considered here. Thus our preliminary analysis indicates that the threat of sanctions does not seem to be an effective way to improve school accountability scores. However a more thorough analysis, including checks for model fit, may come to a different conclusion.

Small study RD estimation methods are important for real data analysis like this, where the data is measured at the school level and there are a relatively small number of schools. There are other scenarios where the treatment is applied at the school level but the data is measured on individual students. The RD methods considered here do not directly account for such clustering, although if the intraclass correlation is low these methods may still be fine to use. See \textcite{Bartalotti2017a} for an example of RD estimation methods that incorporate clustering.

\section*{Discussion}
Regression discontinuity designs will continue to be a popular way to estimate causal effects for educational interventions thanks to the prevalence of cutoff-based programs. Many of the samples used in real world applications are somewhat small or have sparsity around the cutoff, and these characteristics may help determine the type of methodology to use. Rather than using the overall sample size to guide this process, we recommend using our proposed DISS metric, which quantifies the number of observations near the cutoff. In our simulation study and real data example, we showed that $\bar{m}$ was a more useful metric than sample size $n$ in determining whether RD estimates could be obtained from a data set. Furthermore, $\bar{m}$ is useful in understanding trends in the size of calculated bandwidth values, which in turn affects the interval estimate success rates as well as the performance of point and interval estimates. 

We also used our simulation study to compare the performance of popular RD methods when applied to small data sets. Conventional RD inference can lead to problems such as undercoverage, and we saw that for our small study sizes in a variety of scenarios. However, alternative approaches that have been shown to work well for larger sample sizes did not perform markedly better for the small studies under consideration. The robust, bias-corrected approach of CCT requires extra estimation, and if there is not enough data around the cutoff this extra estimation can lead to much higher variability. In our simulation this led to larger MSE values than competing methods, even when paired with the bandwidth algorithm designed for it. Furthermore the coverage of RBC intervals tended to be only slightly higher than conventional intervals while also being quite a bit wider. For study sizes of less than around $m=60$ that are similar to those we studied in the simulation, we do not recommend RBC inference for RD estimation.

The FLCI method proposed by AK improves coverage relative to conventional inference by inflating the critical value. In our simulation we saw this improvement in coverage, but it came at the cost of much wider intervals\added{, particularly for smaller study sizes}. The interval widths stem in part from using the worst-case bias-sd ratio, which can be rather large when samples are small and curvature is large. The AK/FLCI method was competitive in many scenarios in our simulation and may be a good option\added{for study sizes around $m=30$ or greater}, particularly if \added{there is not a lot of curvature in the underlying mean function and }achieving nominal coverage is essential. \deleted{However, this method was also the most likely to fail to produce a finite interval for smaller studies, seemingly because of difficulties with the extra estimation needed to calculate an inflated critical value. Thus it may not be a good method for practitioners with a study size of around $m=30$ or less, and even for some larger study sizes it may be better to stick with a conventional approach if there seems to be a lot of curvature in the underlying mean function.} Recall, however, that we used the data-driven value of $\hat{M}$ in our simulation rather than the true value of $M$. The sensitivity analysis in Section SA2 of the Supplemental Appendix shows that there may be some slight gains from using $M$ instead of $\hat{M}$ in certain situations, but we would expect the number of applications where researchers feel confident using a pre-specified value of $M$ to be small.

Local randomization has a potential advantage over the continuity methods in that it does not rely on large sample approximations and thus might lead to better small study performance. In our simulation the LR method was often competitive and sometimes optimal in terms of MSE for the various DGPs and study sizes, primarily due to small empirical standard errors. However, it was typically the most biased method, and combined with narrow intervals led to very poor coverage in some settings. This bias seems to be worse for more serious violations of the local randomization assumption, which may potentially limit the usefulness of this method in real-world small study applications. Certainly our use of a simple minimum observation window selection limits what we can conclude about local randomization methods for small studies. Perhaps with the right data the window selection algorithm of CFT may improve performance of the LR method for small samples; however such a comparison is beyond the scope of this paper.

In some ways then, our simulation results highlight the difficulty of small study RD estimation. The conventional approach is far from perfect, and in particular leads to lower than desired empirical coverage. But attempts to improve this coverage using the bias correction of CCT, the fixed length confidence intervals of AK, or the finite-sample local randomization methods of CFT show only inconsistent improvement and often come at a cost of other performance measures. The method that paired conventional inference with the IK bandwidth was perhaps the top overall performer across our simulation settings. As we strove to choose a wide variety of simulation settings to examine, this IK/CV combination would be our recommendation for most RD applications with studies below around $m=60$, despite its limitations. Of course, all of our recommendations are delivered with the caveat that patterns we observed could differ for data situations that differ substantially from those we examined by simulation. 

Clearly there is more work to be done in developing RD methodology that works well for small studies. We hope that as researchers take on this task they will do so with a better characterization of the size of an RD study by using the DISS metric $\bar{m}$ in their simulation settings. We believe this will lead to better and more comprehensive guidance for those practitioners analyzing RD data for small studies in the field of education and beyond.

\section*{Acknowledgements}
We would like to thank several anonymous reviewers for their reading of the manuscript and their many helpful suggestions. 

\clearpage
\begin{table}[t!]
\caption{Sample sizes and rule of thumb bandwidths for the different running variable distributions and values of $\bar{m}$, which have been rounded to the nearest whole number.}
\begin{tabularx}{\textwidth}{lIIIIIIIIII} 
\toprule
 & \multicolumn{2}{c}{$\bar{m}=10$} & \multicolumn{2}{c}{$\bar{m}=21$} & \multicolumn{2}{c}{$\bar{m}=27$} & \multicolumn{2}{c}{$\bar{m}=44$} & \multicolumn{2}{c}{$\bar{m}=57$} \\ 
 \cmidrule(lr){2-3} \cmidrule(lr){4-5} \cmidrule(lr){6-7} \cmidrule(lr){8-9} \cmidrule(lr){10-11}
 & $n$ & $h_{ROT}$ & $n$ & $h_{ROT}$ & $n$ & $h_{ROT}$ & $n$ & $h_{ROT}$ & $n$ & $h_{ROT}$\\ 
\midrule
RV1 & 40 & 0.124 & 101 & 0.103 & 140 & 0.097 & 256 & 0.086 & 354 & 0.080 \\ 
RV2 & 56 & 0.072 & 140 & 0.060 & 194 & 0.056 & 354 & 0.050 & 490 & 0.046 \\ 
RV3 & 140 & 0.034 & 354 & 0.028 & 494 & 0.026 & 905 & 0.023 & 1254 & 0.022 \\ 
\bottomrule
\end{tabularx}
\label{mn}
\end{table}

\clearpage
\begin{table}[t!]
\caption{Interval estimate success rates (in percents) for estimation methods at various study sizes for RV2$\mu_2$.}
\begin{tabularx}{\textwidth}{EIIIIIIIIII}
\toprule
\multicolumn{1}{l}{} & \multicolumn{3}{c}{RBC} & \multicolumn{3}{c}{CV} & \multicolumn{3}{c}{FLCI} &  \\ 
\cmidrule(lr){2-4} \cmidrule(lr){5-7} \cmidrule(lr){8-10}
\multicolumn{1}{l}{$\bar{m}$} & IK & AK & CCT & IK & AK & CCT & IK & AK & CCT & LR5 \\ 
\midrule
10 & $97.37$ & $66.41$ & $65.57$ & $97.37$ & $67.11$ & $65.59$ & $97.70$ & $98.57$ & $65.46$ & $100.00$ \\ 
21 & $100.00$ & $97.07$ & $99.45$ & $100.00$ & $97.09$ & $99.45$ & $100.00$ & $100.00$ & $99.44$ & $100.00$ \\ 
27 & $100.00$ & $99.35$ & $99.95$ & $100.00$ & $99.36$ & $99.95$ & $100.00$ & $100.00$ & $99.95$ & $100.00$ \\ 
44 & $100.00$ & $99.99$ & $100.00$ & $100.00$ & $99.99$ & $100.00$ & $100.00$ & $100.00$ & $100.00$ & $100.00$ \\ 
57 & $100.00$ & $100.00$ & $100.00$ & $100.00$ & $100.00$ & $100.00$ & $100.00$ & $100.00$ & $100.00$ & $100.00$ \\ 
\bottomrule
\end{tabularx}
\label{tab1}
\end{table}

\clearpage
\begin{table}[t!]
\caption{Overall sample size, sample size below the cutoff, Silverman rule of thumb bandwidth, and number of observations within $h_{ROT}$ of the cutoff for selected subsets of Indiana schools.}
\begin{tabularx}{\textwidth}{lrrII}
\toprule
Subset & Sample Size & Sample Size Below & $h_{ROT}(s^*)$ & $m$\\ 
\midrule
All & 1933 & 88 & 2.31 & 51\\ 
Low & 1676 & 86 & 2.42 & 51\\ 
Traditional Public & 1625 & 77 & 2.30 & 42\\ 
High & 408 & 7 & 1.88 & 4\\ 
Private & 308 & 11 & 3.60 & 10\\ 
\bottomrule
\end{tabularx}
\label{subset}
\end{table}

\clearpage

\clearpage

\clearpage

\clearpage

\clearpage

\clearpage

\clearpage

\clearpage

\clearpage
\printbibliography

\newpage
\setcounter{page}{1}


\setcounter{figure}{0} 
\setcounter{table}{0}

{Supplemental Appendix to ``Small Study Regression Discontinuity Designs: Density Inclusive Study Size Metric and Performance"}

\maketitle

\setcounter{section}{0}
\renewcommand{\thefigure}{SA\arabic{figure}}
\renewcommand{\thetable}{SA\arabic{table}}
\renewcommand{\thesection}{SA\arabic{section}}

\newpage
\section{Additional Tables and Figures}

\begin{table}[htb!]
\caption{Bandwidth success rates for all DGPs. Bandwidth algorithms were universally successful at study sizes larger than those listed.}
\setlength{\tabcolsep}{3pt}
\begin{tabularx}{\textwidth}{l|IIIIIIIII}
\toprule
\multicolumn{1}{l}{} & \multicolumn{3}{c}{IK} & \multicolumn{3}{c}{AK} & \multicolumn{3}{c}{CCT} \\ 
\cmidrule(lr){2-4} \cmidrule(lr){5-7} \cmidrule(lr){8-10}
\multicolumn{1}{l}{$\bar{m}$} & $\mu_1$ & $\mu_2$ & $\mu_3$ & $\mu_1$ & $\mu_2$ & $\mu_3$ & $\mu_1$ & $\mu_2$ & $\mu_3$ \\ 
\midrule
\multicolumn{1}{l}{RV1} \\ 
\midrule
10 & $99.98$ & $99.82$ & $99.97$ & $100.00$ & $99.39$ & $100.00$ & $96.94$ & $98.36$ & $96.35$ \\ 
21 & $100.00$ & $100.00$ & $100.00$ & $100.00$ & $100.00$ & $100.00$ & $99.99$ & $100.00$ & $100.00$ \\ 
27 & $100.00$ & $100.00$ & $100.00$ & $100.00$ & $100.00$ & $100.00$ & $100.00$ & $100.00$ & $100.00$ \\ 
\midrule
\multicolumn{1}{l}{RV2} \\ 
\midrule
10 &99.43&	99.21&	99.42&	98.71	&98.57	&98.71	&82.79&	82.69	&82.38\\
21 & $100.00$ & $100.00$ & $100.00$ & $100.00$ & $100.00$ & $100.00$ & $99.98$ & $99.97$ & $99.97$ \\ 
27 & $100.00$ & $100.00$ & $100.00$ & $100.00$ & $100.00$ & $100.00$ & $100.00$ & $100.00$ & $100.00$ \\ 
\midrule
\multicolumn{1}{l}{RV3} \\ 
\midrule
10 & 97.89&	97.90&	97.90&	91.03&	91.03	&91.03&	64.50&	64.58&	64.23 \\ 
21 & $100.00$ & $100.00$ & $100.00$ & $100.00$ & $100.00$ & $100.00$ & $99.79$ & $99.78$ & $99.79$ \\ 
27 & $100.00$ & $100.00$ & $100.00$ & $100.00$ & $100.00$ & $100.00$ & $100.00$ & $100.00$ & $100.00$ \\ 
\bottomrule
\end{tabularx}
\label{tabbw}
\end{table}

\begin{table}[htb!]
\caption{Interval estimate success rates (in percents) for $\mu_1$.}
\setlength{\tabcolsep}{3pt}
\begin{tabularx}{\textwidth}{llIIIIIIIIII}
\toprule
 &  & \multicolumn{3}{c}{RBC} & \multicolumn{3}{c}{CV} & \multicolumn{3}{c}{FLCI} &  \\ 
\cmidrule(lr){3-5} \cmidrule(lr){6-8} \cmidrule(lr){9-11}
$\bar{m}$ & n & IK & AK & CCT & IK & AK & CCT & IK & AK & CCT & LR5 \\ 
\midrule
\multicolumn{1}{l}{RV1} \\ 
\midrule
10 & 40 & $99.83$ & $89.93$ & $82.58$ & $99.83$ & $90.06$ & $82.56$ & $99.88$ & $100.00$ & $82.49$ & $100.00$ \\ 
21 & 101 & $100.00$ & $99.96$ & $99.76$ & $100.00$ & $99.96$ & $99.76$ & $100.00$ & $100.00$ & $99.76$ & $100.00$ \\ 
27 & 140 & $100.00$ & $100.00$ & $99.98$ & $100.00$ & $100.00$ & $99.98$ & $100.00$ & $100.00$ & $99.98$ & $100.00$ \\ 
44 & 256 & $100.00$ & $100.00$ & $100.00$ & $100.00$ & $100.00$ & $100.00$ & $100.00$ & $100.00$ & $100.00$ & $100.00$ \\ 
57 & 354 & $100.00$ & $100.00$ & $100.00$ & $100.00$ & $100.00$ & $100.00$ & $100.00$ & $100.00$ & $100.00$ & $100.00$ \\ 
\midrule
\multicolumn{1}{l}{RV2} \\ 
\midrule
10 & 56 & $98.89$ & $80.47$ & $66.41$ & $98.89$ & $80.87$ & $66.42$ & $98.32$ & $98.71$ & $66.30$ & $100.00$ \\ 
21 & 140 & $100.00$ & $99.85$ & $99.49$ & $100.00$ & $99.85$ & $99.49$ & $100.00$ & $100.00$ & $99.49$ & $100.00$ \\ 
27 & 194 & $100.00$ & $99.99$ & $99.97$ & $100.00$ & $99.99$ & $99.96$ & $100.00$ & $100.00$ & $99.96$ & $100.00$ \\ 
44 & 354 & $100.00$ & $100.00$ & $100.00$ & $100.00$ & $100.00$ & $100.00$ & $100.00$ & $100.00$ & $100.00$ & $100.00$ \\ 
57 & 490 & $100.00$ & $100.00$ & $100.00$ & $100.00$ & $100.00$ & $100.00$ & $100.00$ & $100.00$ & $100.00$ & $100.00$ \\
\midrule
\multicolumn{1}{l}{RV3} \\ 
\midrule
10 & 140 & $95.72$ & $64.67$ & $48.97$ & $95.72$ & $65.28$ & $48.97$ & $90.66$ & $91.03$ & $48.78$ & $99.97$ \\ 
21 & 354 & $100.00$ & $99.25$ & $98.51$ & $100.00$ & $99.25$ & $98.51$ & $99.99$ & $100.00$ & $98.49$ & $100.00$ \\ 
27 & 494 & $100.00$ & $99.94$ & $99.87$ & $100.00$ & $99.94$ & $99.87$ & $100.00$ & $100.00$ & $99.86$ & $100.00$ \\ 
44 & 905 & $100.00$ & $100.00$ & $100.00$ & $100.00$ & $100.00$ & $100.00$ & $100.00$ & $100.00$ & $100.00$ & $100.00$ \\ 
57 & 1254 & $100.00$ & $100.00$ & $100.00$ & $100.00$ & $100.00$ & $100.00$ & $100.00$ & $100.00$ & $100.00$ & $100.00$ \\ 
\bottomrule
\end{tabularx}
\label{tab2}
\end{table}

\begin{table}
\caption{Interval estimate success rates (in percents) for $\mu_2$.}
\setlength{\tabcolsep}{3pt}
\begin{tabularx}{\textwidth}{llIIIIIIIIII}
\toprule
 &  & \multicolumn{3}{c}{RBC} & \multicolumn{3}{c}{CV} & \multicolumn{3}{c}{FLCI} &  \\ 
\cmidrule(lr){3-5} \cmidrule(lr){6-8} \cmidrule(lr){9-11}
$\bar{m}$ & n & IK & AK & CCT & IK & AK & CCT & IK & AK & CCT & LR5 \\ 
\midrule
\multicolumn{1}{l}{RV1} \\ 
\midrule
10 & 40 & $97.16$ & $52.51$ & $82.68$ & $97.16$ & $53.62$ & $82.67$ & $98.68$ & $99.39$ & $82.59$ & $100.00$ \\ 
21 & 101 & $99.98$ & $84.68$ & $99.05$ & $99.98$ & $84.84$ & $99.03$ & $100.00$ & $100.00$ & $99.02$ & $100.00$ \\ 
27 & 140 & $100.00$ & $93.28$ & $99.86$ & $100.00$ & $93.33$ & $99.87$ & $100.00$ & $100.00$ & $99.86$ & $100.00$ \\ 
44 & 256 & $100.00$ & $99.61$ & $100.00$ & $100.00$ & $99.61$ & $100.00$ & $100.00$ & $100.00$ & $100.00$ & $100.00$ \\ 
57 & 354 & $100.00$ & $99.96$ & $100.00$ & $100.00$ & $99.96$ & $100.00$ & $100.00$ & $100.00$ & $100.00$ & $100.00$ \\ 
\midrule
\multicolumn{1}{l}{RV2} \\ 
\midrule
10 & 56 & $97.37$ & $66.41$ & $65.57$ & $97.37$ & $67.11$ & $65.59$ & $97.70$ & $98.57$ & $65.46$ & $100.00$ \\ 
21 & 140 & $100.00$ & $97.07$ & $99.45$ & $100.00$ & $97.09$ & $99.45$ & $100.00$ & $100.00$ & $99.44$ & $100.00$ \\ 
27 & 194 & $100.00$ & $99.35$ & $99.95$ & $100.00$ & $99.36$ & $99.95$ & $100.00$ & $100.00$ & $99.95$ & $100.00$ \\ 
44 & 354 & $100.00$ & $99.99$ & $100.00$ & $100.00$ & $99.99$ & $100.00$ & $100.00$ & $100.00$ & $100.00$ & $100.00$ \\ 
57 & 490 & $100.00$ & $100.00$ & $100.00$ & $100.00$ & $100.00$ & $100.00$ & $100.00$ & $100.00$ & $100.00$ & $100.00$ \\ 
\midrule
\multicolumn{1}{l}{RV3} \\ 
\midrule
10 & 140 & $95.57$ & $64.33$ & $49.09$ & $95.57$ & $64.91$ & $49.08$ & $90.63$ & $91.03$ & $48.88$ & $99.97$ \\ 
21 & 354 & $100.00$ & $99.27$ & $98.50$ & $100.00$ & $99.28$ & $98.49$ & $99.99$ & $100.00$ & $98.48$ & $100.00$ \\ 
27 & 494 & $100.00$ & $99.96$ & $99.86$ & $100.00$ & $99.96$ & $99.86$ & $100.00$ & $100.00$ & $99.86$ & $100.00$ \\ 
44 & 905 & $100.00$ & $100.00$ & $100.00$ & $100.00$ & $100.00$ & $100.00$ & $100.00$ & $100.00$ & $100.00$ & $100.00$ \\ 
57 & 1254 & $100.00$ & $100.00$ & $100.00$ & $100.00$ & $100.00$ & $100.00$ & $100.00$ & $100.00$ & $100.00$ & $100.00$ \\
\bottomrule
\end{tabularx}
\label{tab3}
\end{table}

\begin{table}
\caption{Interval estimate success rates (in percents) for $\mu_3$.}
\setlength{\tabcolsep}{3pt}
\begin{tabularx}{\textwidth}{llIIIIIIIIII}
\toprule
 &  & \multicolumn{3}{c}{RBC} & \multicolumn{3}{c}{CV} & \multicolumn{3}{c}{FLCI} &  \\ 
\cmidrule(lr){3-5} \cmidrule(lr){6-8} \cmidrule(lr){9-11}
$\bar{m}$ & n & IK & AK & CCT & IK & AK & CCT & IK & AK & CCT & LR5 \\ 
\midrule
\multicolumn{1}{l}{RV1} \\ 
\midrule
10 & 40 & $99.59$ & $88.55$ & $80.31$ & $99.59$ & $88.70$ & $80.29$ & $99.76$ & $100.00$ & $80.21$ & $100.00$ \\ 
21 & 101 & $100.00$ & $99.94$ & $99.77$ & $100.00$ & $99.94$ & $99.77$ & $100.00$ & $100.00$ & $99.77$ & $100.00$ \\ 
27 & 140 & $100.00$ & $100.00$ & $99.98$ & $100.00$ & $100.00$ & $99.98$ & $100.00$ & $100.00$ & $99.98$ & $100.00$ \\ 
44 & 256 & $100.00$ & $100.00$ & $100.00$ & $100.00$ & $100.00$ & $100.00$ & $100.00$ & $100.00$ & $100.00$ & $100.00$ \\ 
57 & 354 & $100.00$ & $100.00$ & $100.00$ & $100.00$ & $100.00$ & $100.00$ & $100.00$ & $100.00$ & $100.00$ & $100.00$ \\ 
\midrule
\multicolumn{1}{l}{RV2} \\ 
\midrule
10 & 56 & $98.74$ & $80.23$ & $65.40$ & $98.74$ & $80.66$ & $65.42$ & $98.28$ & $98.71$ & $65.29$ & $100.00$ \\ 
21 & 140 & $100.00$ & $99.86$ & $99.49$ & $100.00$ & $99.86$ & $99.50$ & $100.00$ & $100.00$ & $99.49$ & $100.00$ \\ 
27 & 194 & $100.00$ & $100.00$ & $99.96$ & $100.00$ & $100.00$ & $99.96$ & $100.00$ & $100.00$ & $99.96$ & $100.00$ \\ 
44 & 354 & $100.00$ & $100.00$ & $100.00$ & $100.00$ & $100.00$ & $100.00$ & $100.00$ & $100.00$ & $100.00$ & $100.00$ \\ 
57 & 490 & $100.00$ & $100.00$ & $100.00$ & $100.00$ & $100.00$ & $100.00$ & $100.00$ & $100.00$ & $100.00$ & $100.00$ \\ 
\midrule
\multicolumn{1}{l}{RV3} \\ 
\midrule
10 & 140 & $95.71$ & $64.64$ & $48.65$ & $95.71$ & $65.25$ & $48.65$ & $90.66$ & $91.03$ & $48.47$ & $99.97$ \\ 
21 & 354 & $99.99$ & $99.30$ & $98.49$ & $99.99$ & $99.31$ & $98.50$ & $99.99$ & $100.00$ & $98.49$ & $100.00$ \\ 
27 & 494 & $100.00$ & $99.96$ & $99.87$ & $100.00$ & $99.96$ & $99.87$ & $100.00$ & $100.00$ & $99.86$ & $100.00$ \\ 
44 & 905 & $100.00$ & $100.00$ & $100.00$ & $100.00$ & $100.00$ & $100.00$ & $100.00$ & $100.00$ & $100.00$ & $100.00$ \\ 
57 & 1254 & $100.00$ & $100.00$ & $100.00$ & $100.00$ & $100.00$ & $100.00$ & $100.00$ & $100.00$ & $100.00$ & $100.00$ \\ 
\bottomrule
\end{tabularx}
\label{tab4}
\end{table}

\begin{figure}[t!]
    \centering
    \includegraphics[width=\textwidth]{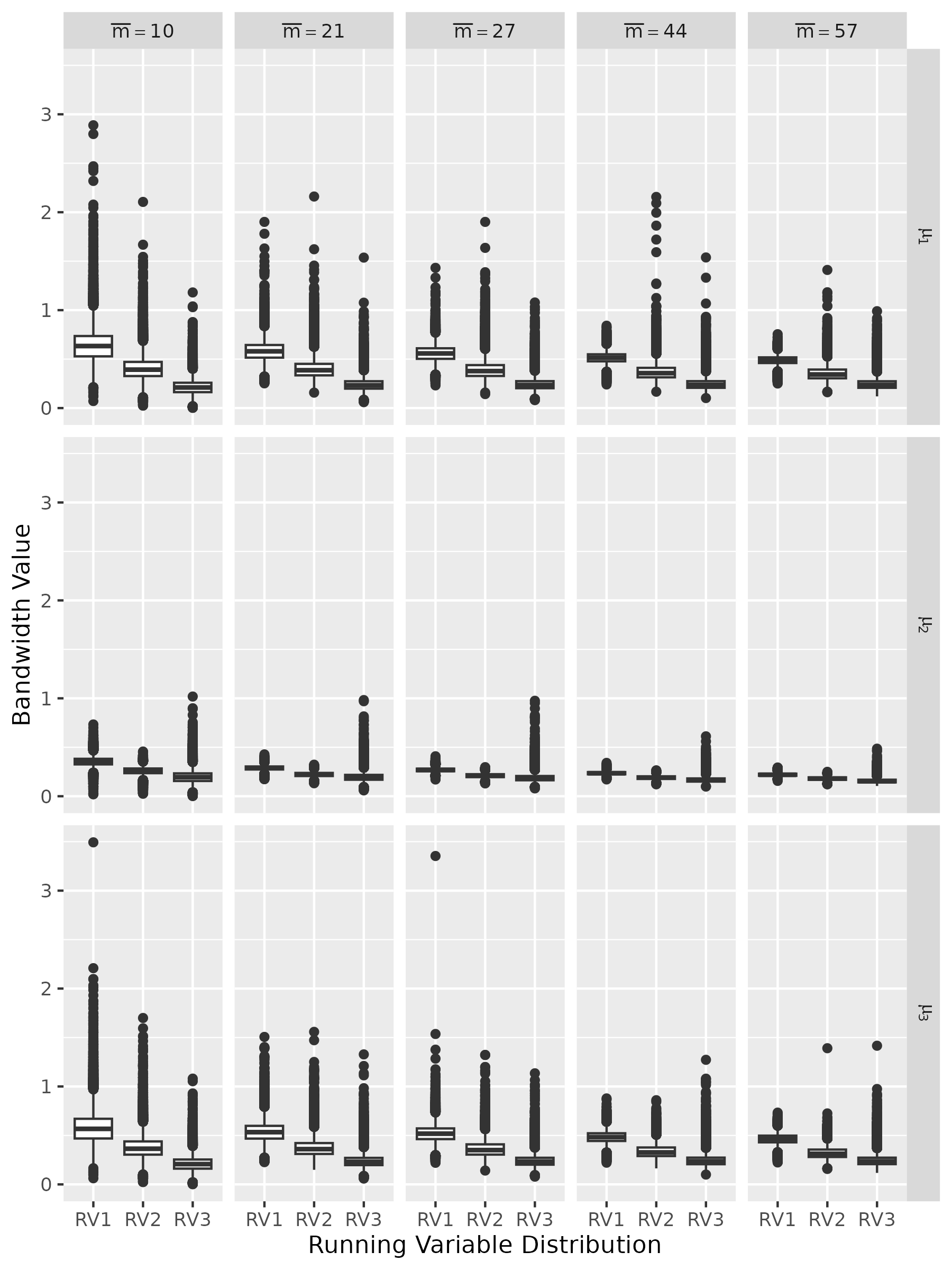}
    \caption{Bandwidth distributions for the IK algorithm.}
    \label{bwik1}
\end{figure}

\begin{figure}[t!]
    \centering
    \includegraphics[width=\textwidth]{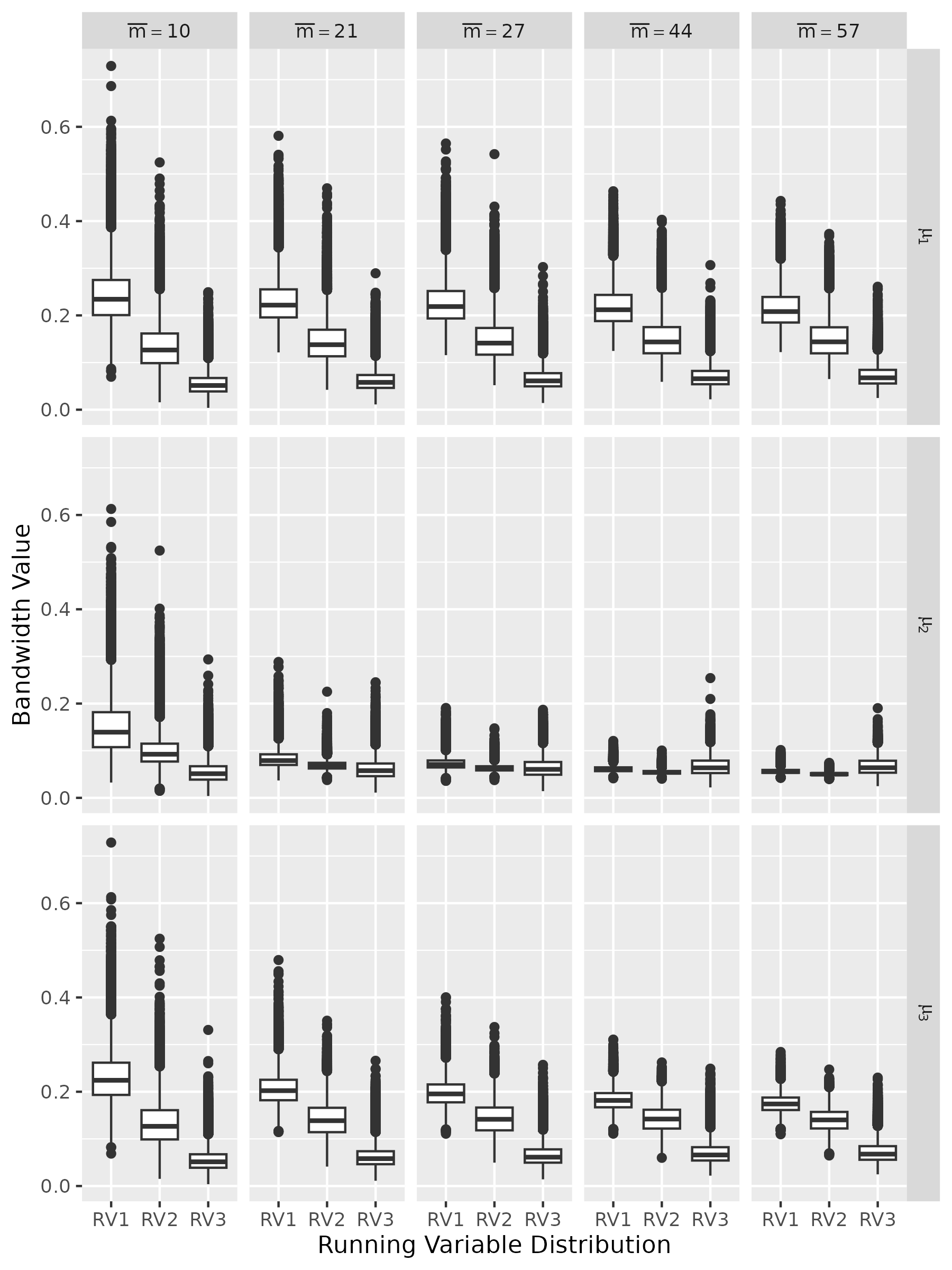}
    \caption{Bandwidth distributions for the AK algorithm.}
    \label{bwak1}
\end{figure}

\begin{figure}[t!]
    \centering
    \includegraphics[width=\textwidth]{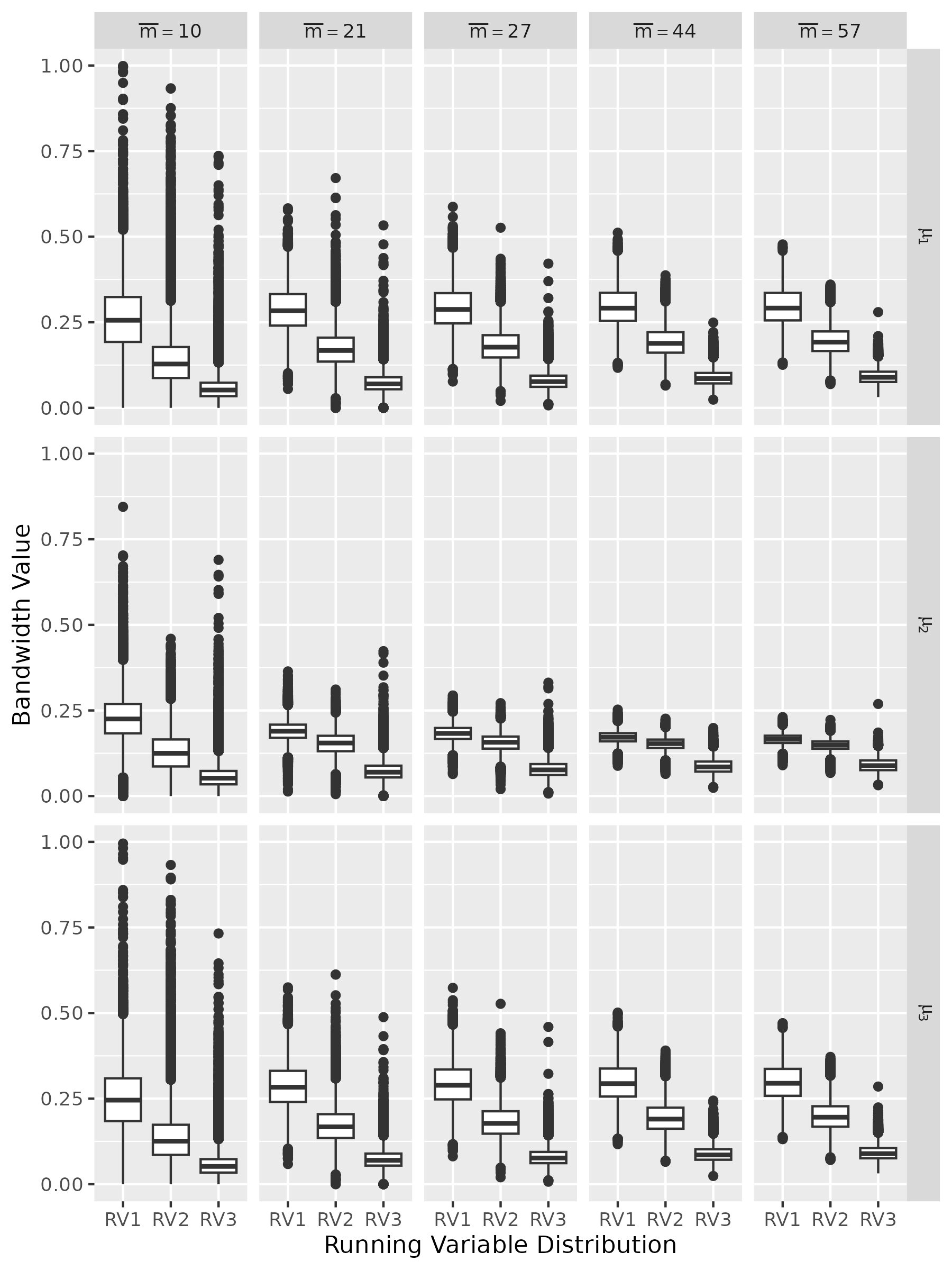}
    \caption{Bandwidth distributions for the CCT algorithm.}
    \label{bwcct1}
\end{figure}

\begin{figure}[t!]
    \centering
    \includegraphics[width=\textwidth]{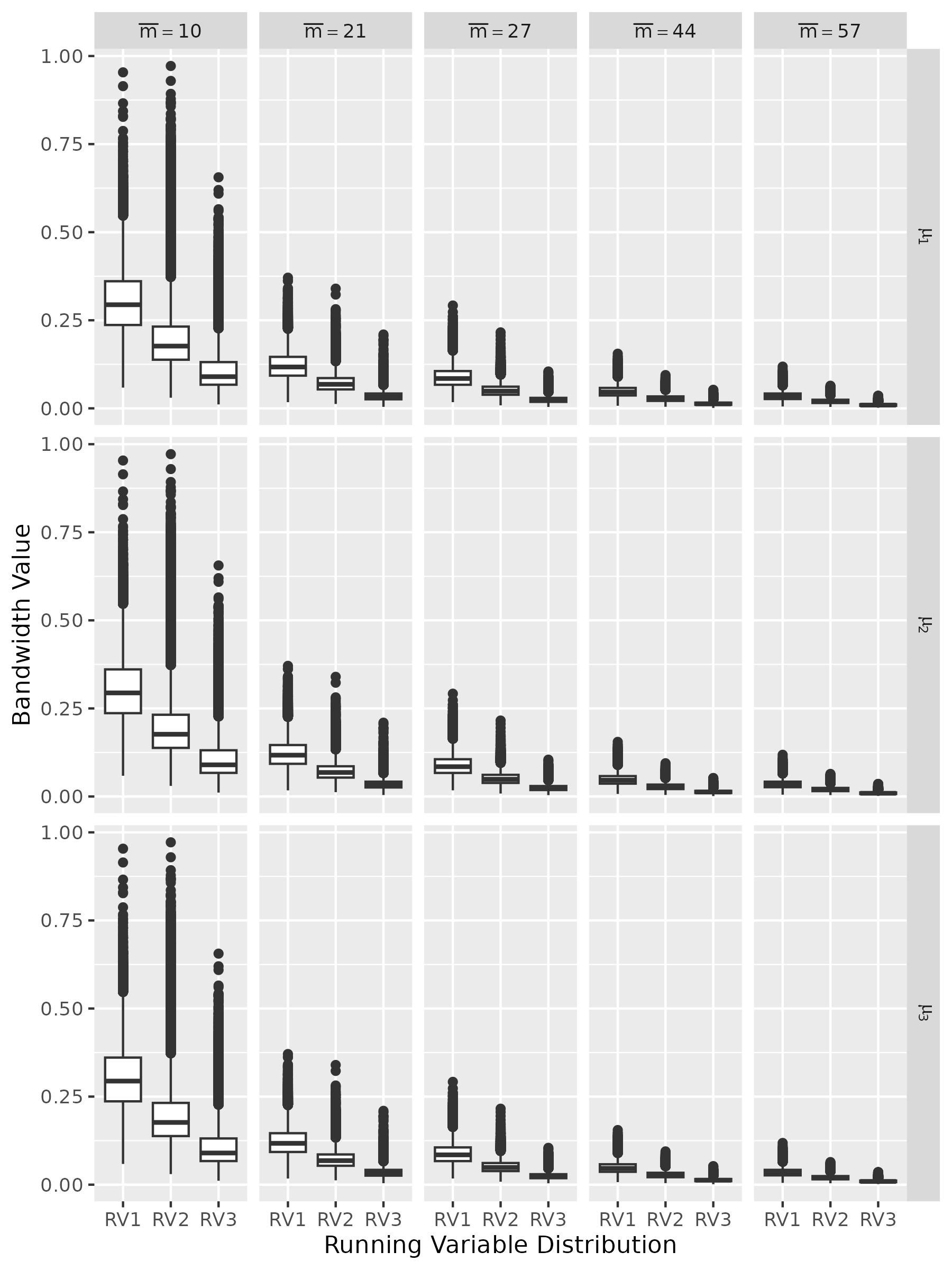}
    \caption{Window distributions for the LR5 algorithm.}
    \label{bwlr5}
\end{figure}

\begin{figure}[t!]
    \centering
    \includegraphics[width=\textwidth]{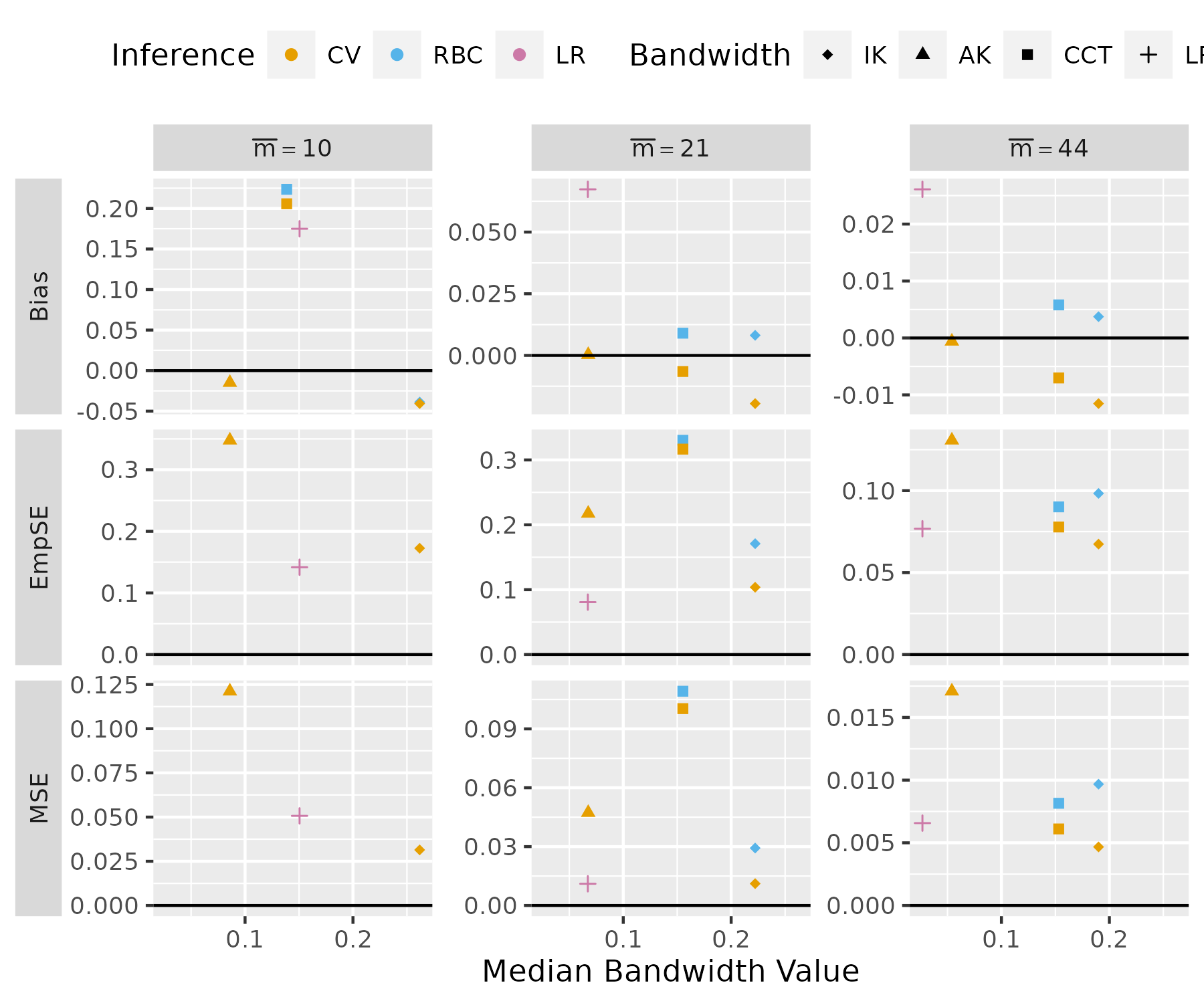}
    \caption{Performance measures for methods using CV, RBC, and LR inference for RV2$\mu_2$ at three study sizes. \textit{FLCI values are essentially the same as CV and thus omitted. Values based on iterations in which all estimates where finite, 46.7$\%$ of all iterations for $\bar{m}=10$, 97$\%$ for $\bar{m}=21$, and 99.99$\%$ for $\bar{m}=44$. The RBC/AK method is not competitive in terms of MSE for any study size and thus omitted. For $\bar{m}=10$ the graph also omits EmpSE and MSE values for CV/CCT and RBC/CCT (28 and 791, respectively) and RBC/IK (6 and 35). The maximum Monte Carlo standard errors (MCSE) for the values shown for $\bar{m}=10$ are 0.005. The maximum MCSE for the values shown for $\bar{m}=21$ are 0.002 except for MSE for RBC/CCT and CV/CCT (0.02). The maximum MCSE for the values shown for $\bar{m}=44$ are 0.0006.}}
    \label{lrscatter3}
\end{figure}

\begin{figure}[t!]
    \centering
    \includegraphics[width=\textwidth]{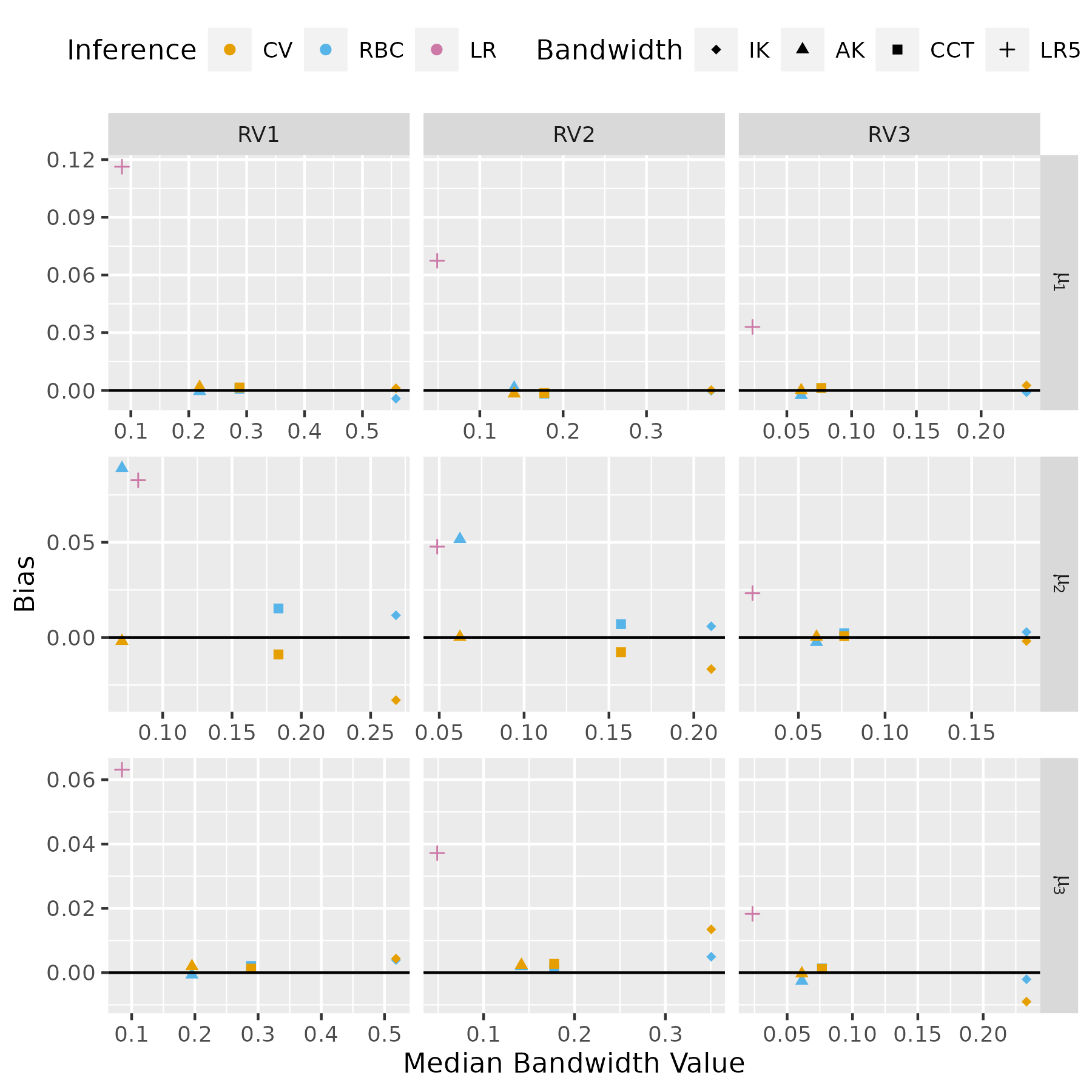}
    \caption{Bias of point estimates for different methods and designs at $\bar{m}=27$. \textit{FLCI values are essentially the same as CV and thus omitted. Values based on iterations in which all estimates were finite, which was at least 99$\%$ of iterations except for RV1$\mu_2$ (93$\%$). All MCSE are at most 0.002 except AK/RBC for RV1$\mu_2$ (.09) and RV2$\mu_2$ (.06).}}
    \label{point.scatter.bias}
\end{figure}

\begin{figure}[t!]
    \centering
    \includegraphics[width=\textwidth]{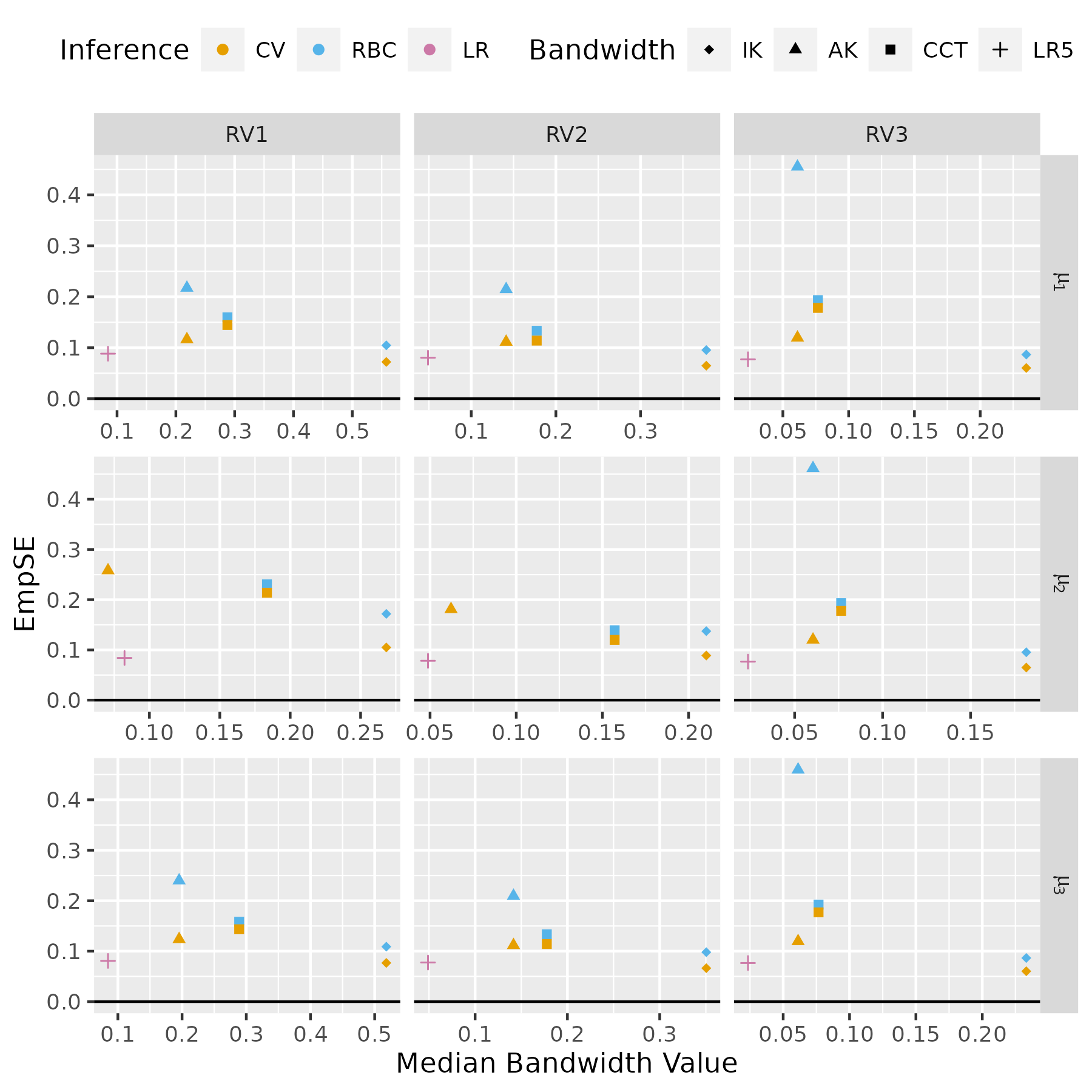}
    \caption{Empirical standard errors of point estimates for different methods and designs at $\bar{m}=27$. \textit{FLCI values are essentially the same as CV and thus omitted. Values based on iterations in which all estimates were finite, which was at least 99$\%$ of iterations except for RV1$\mu_2$ (93$\%$). The graph omits the value of AK/RBC for RV1$\mu_2$ (20) and RV2$\mu_2$ (14). All estimates shown have MCSE at most 0.001.}}
    \label{point.scatter.empse}
\end{figure}

\begin{figure}[t!]
    \centering
    \includegraphics[width=\textwidth]{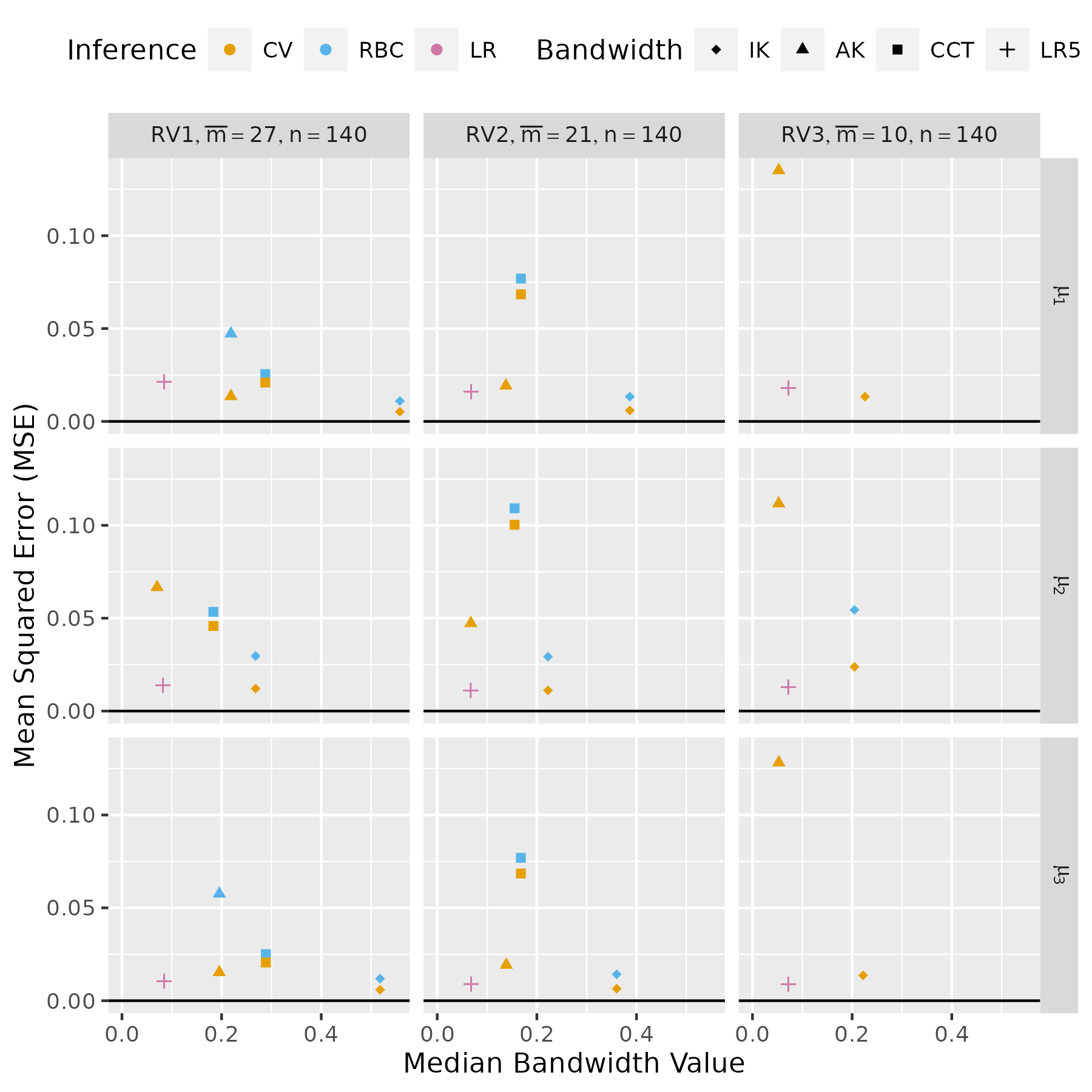}
    \caption{MSE of point estimates and median bandwidth values for three DGPs with overall sample size of $n=140$. \textit{FLCI values are essentially the same as CV and thus omitted. Values based on iterations in which all estimates were finite, which was at least 93$\%$ for the cells in the left panel,  at least 96$\%$ for the cells in the middle panel, and around 39$\%$ for the right panel. The panel on the left omits the AK/RBC value of 410 ($\mu_2$). The center panel omits AK/RBC values of 5 ($\mu_1$), 238 ($\mu_2$), and 5 ($\mu_3$). The panel on the right omits AK/RBC values of around $8.5\cdot10^6$ for each $\mu$, CCT/RBC and CCT/CV values of around 6000 for each $\mu$, and IK/RBC values of 0.39 ($\mu_1$) and 0.31 ($\mu_3$). All estimates shown in the left panel have MCSE at most 0.01. All estimates shown in the center panels have MCSE at most 0.001 except those with CCT, which have a maximum of 0.02. All estimates shown in the right panel have MCSE at most 0.02.}}
    \label{lrscattercomp}
\end{figure}

\begin{figure}[t!]
    \centering
    \includegraphics[width=\textwidth]{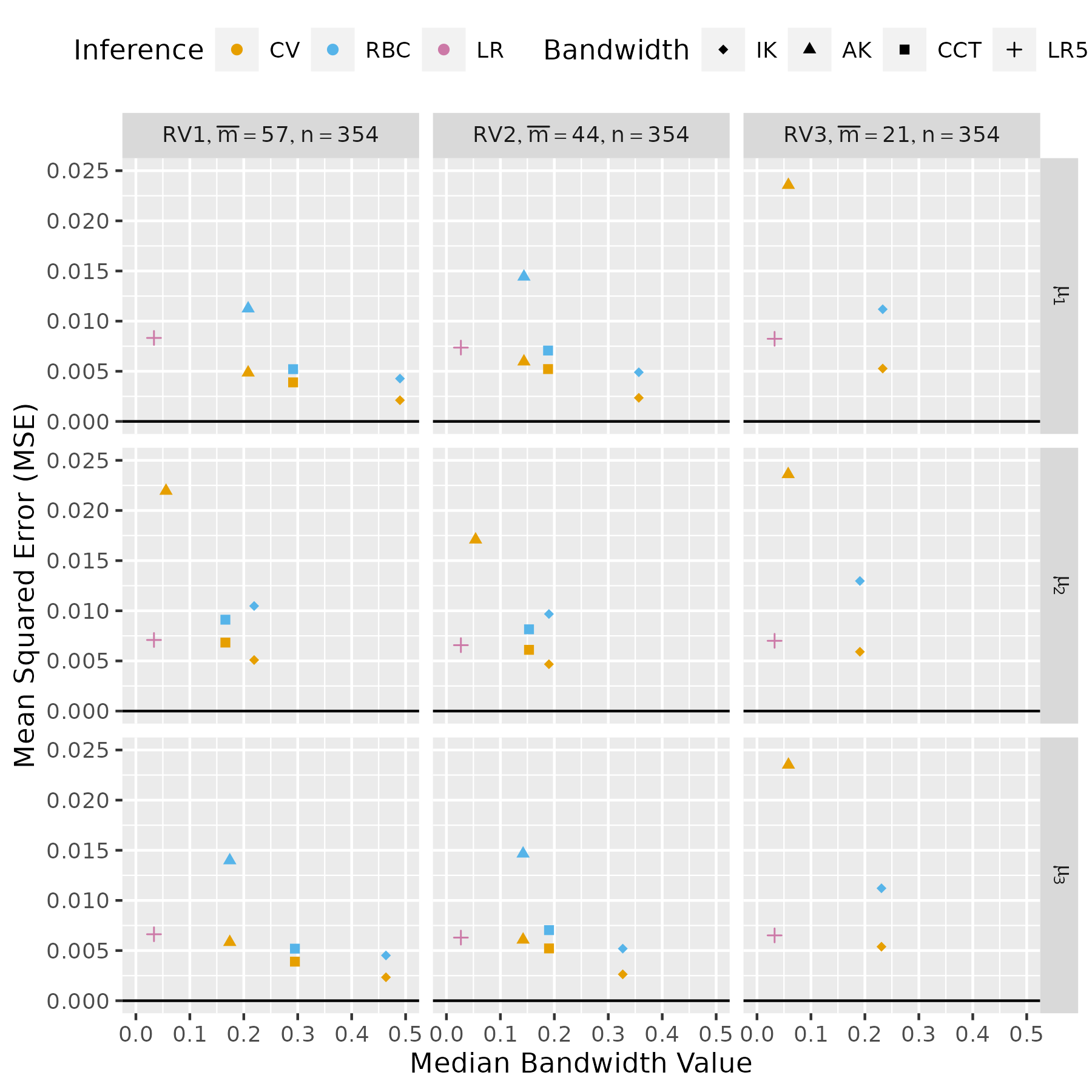}
    \caption{MSE of point estimates and median bandwidth values for three DGPs with overall sample size of $n=354$. \textit{FLCI values are essentially the same as CV and thus omitted. The panel on the left omits the AK/RBC value of 0.19 ($\mu_2$). The center panel omits the AK/RBC value of 0.08 ($\mu_2$). The panel on the right omits AK/RBC values of around 17 ($\mu_1$, $\mu_2$, and $\mu_3$) and values of all methods with CCT of around 0.8 ($\mu_1$, $\mu_2$, and $\mu_3$). All estimates shown have MCSE at most 0.0002.}}
    \label{lrscattercomp2}
\end{figure}

\begin{figure}[t!]
    \centering
    \includegraphics[width=\textwidth]{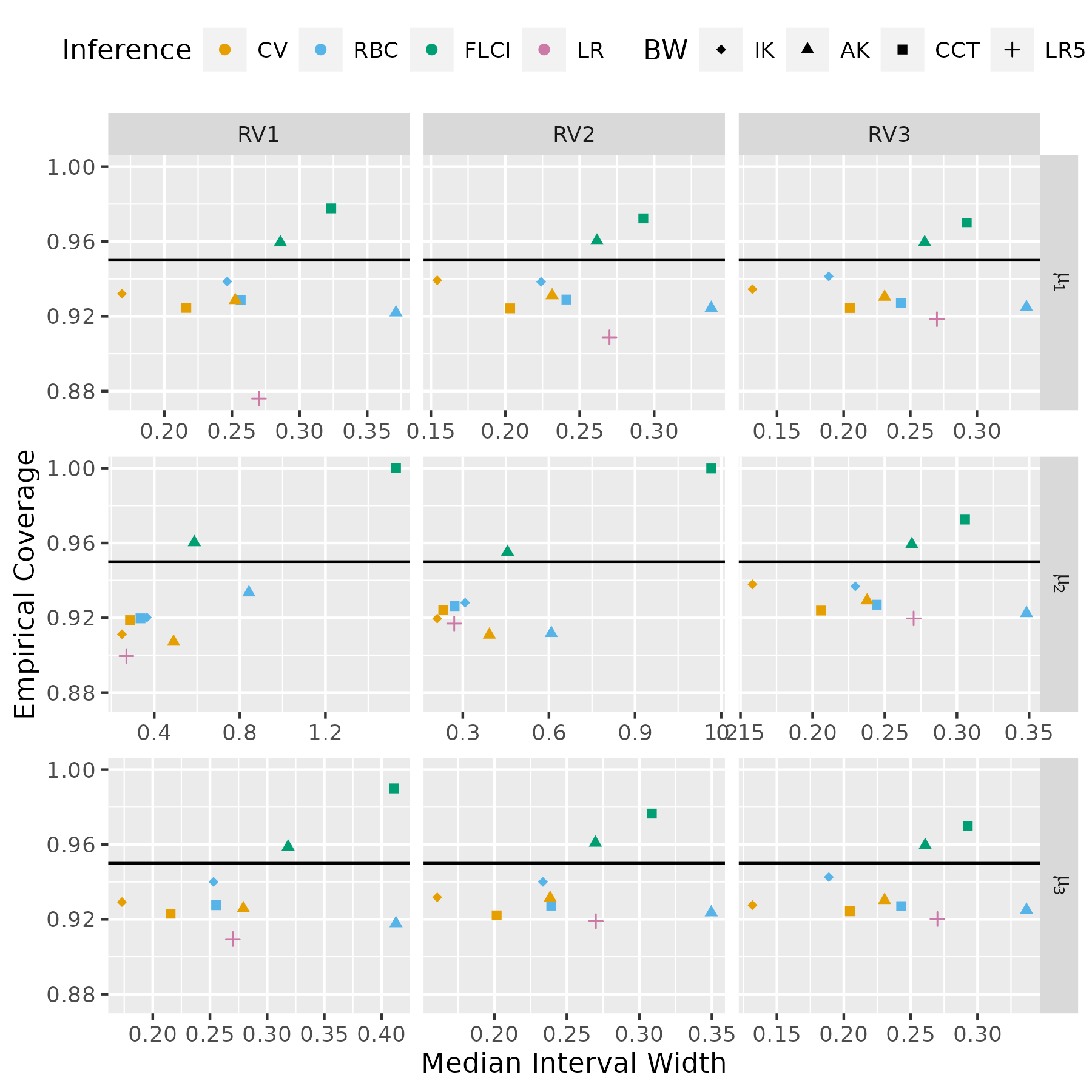}
    \caption{Empirical coverage and median interval widths at $\bar{m}=57$ for all DGPs. \textit{Values based on iterations in which all estimates were finite, which are at 100$\%$ except for RV1$\mu_2$ (99.96$\%$). The graph omits values of IK/FLCI. The maximum Monte Carlo standard error for the coverage estimates is 0.001.}}
    \label{coverage3}
\end{figure}

\begin{figure}[t!]
    \centering
    \includegraphics[width=\textwidth]{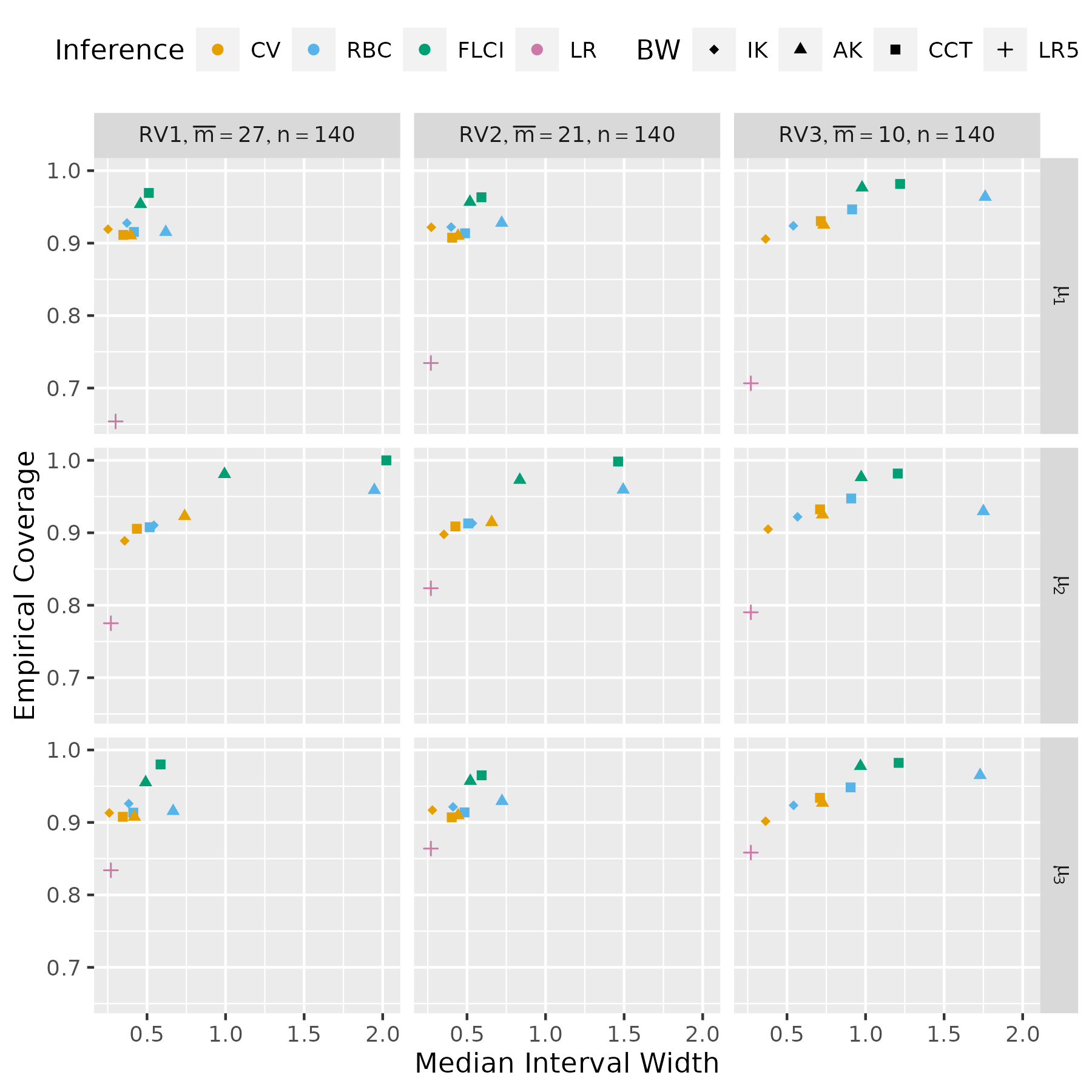}
    \caption{Empirical coverage and median interval width for three DGPs with overall sample size of $n=140$. \textit{Values based on iterations in which all estimates were finite, which was at least 93$\%$ for the cells in the left panel,  at least 96$\%$ for the cells in the middle panel, and around 39$\%$ for the right panel. The graph omits values for IK/FLCI. All coverage estimates shown have MCSE at most 0.003.}}
    \label{lrcvcomp}
\end{figure}

\begin{figure}[t!]
    \centering
    \includegraphics[width=\textwidth]{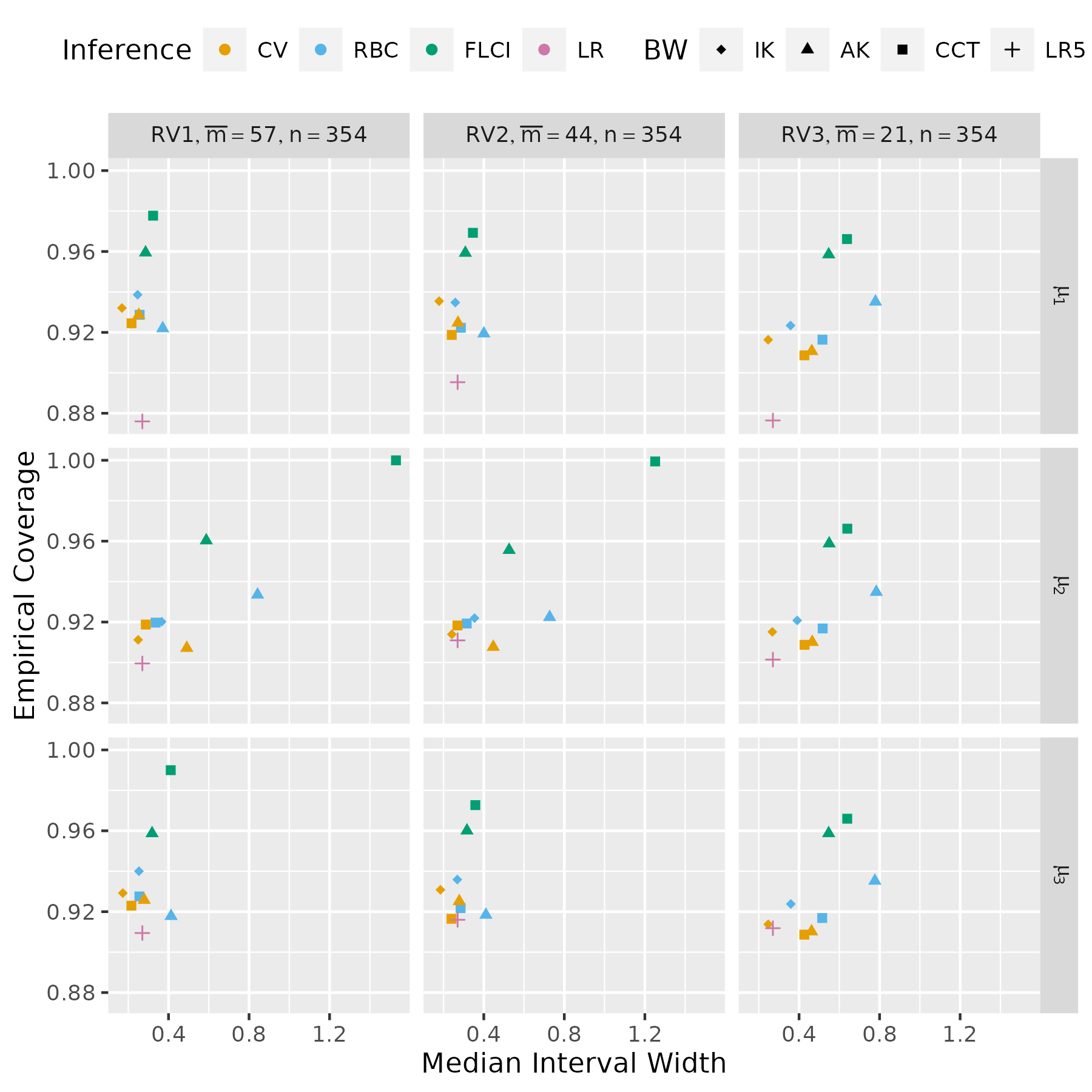}
    \caption{Empirical coverage and median interval width for three DGPs with overall sample size of $n=354$. \textit{The graph omits values for IK/FLCI. All coverage estimates shown have MCSE at most 0.001.}}
    \label{lrcvcomp2}
\end{figure}

\begin{figure}[t!]
    \centering
    \includegraphics[width=\textwidth]{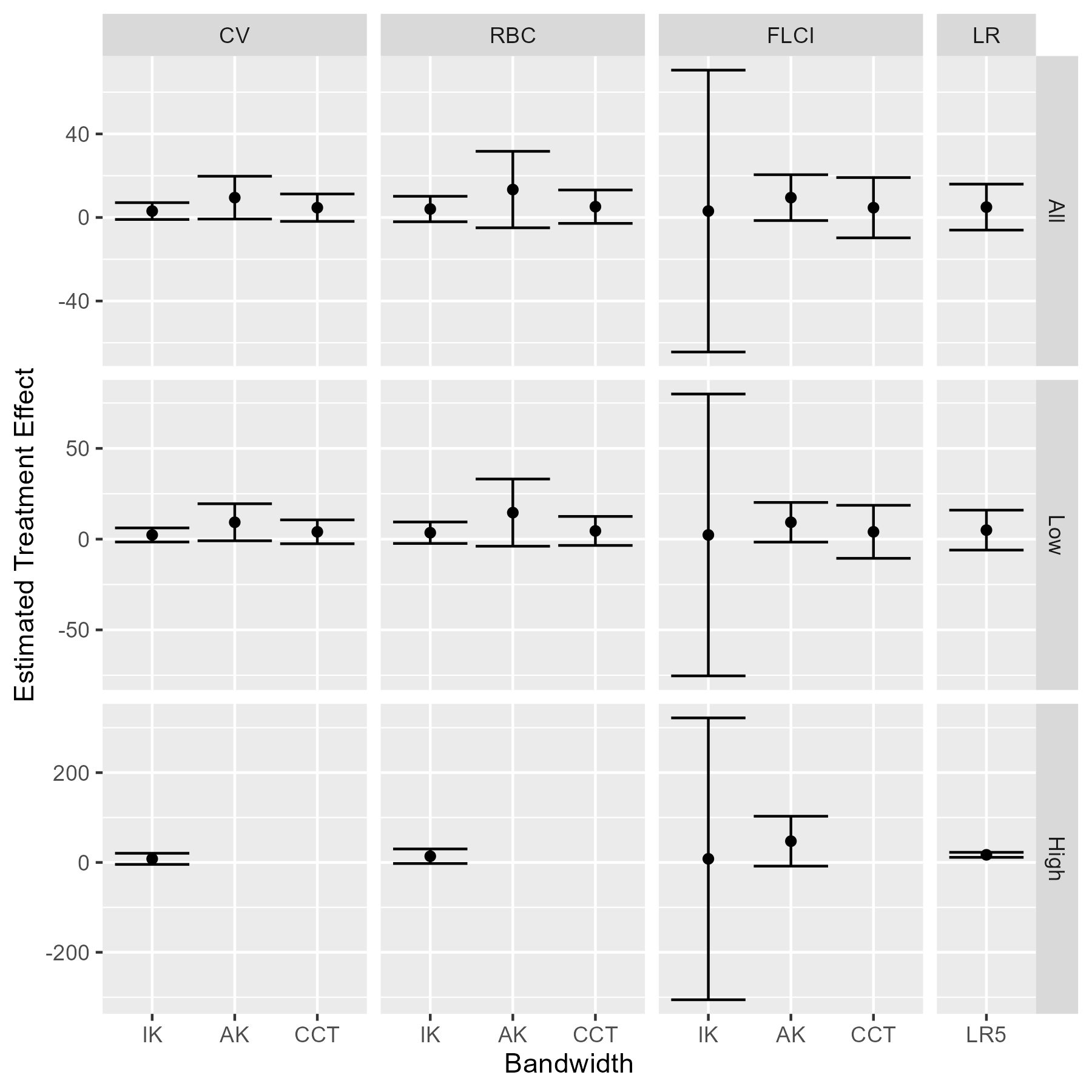}
    \caption{Treatment effect estimates and 90$\%$ confidence intervals for select Indiana subsets.}
    \label{indiana.ci2}
\end{figure}

\clearpage
\begin{refsection}
\section{Sensitivity Analysis} \label{sensanal}
In our main simulation results we used the data driven $\hat{M}$ for the second derivative bound in the AK and FLCI methods. However, \textcite{armstrong2020simple} argue that the choice of $M$ should be made a priori in order to maintain the honesty of their intervals. Clearly knowledge of the true value of $M$ is an advantage as it allows estimation of fewer unknown quantities. To determine how much of an improvement this knowledge provides we \added{present simulation results here that} include methods \deleted{in our simulation }that make use of this true value of $M$. We let AKM\deleted{1, AKM2,} and FLCIM refer to the algorithm\deleted{s} and technique\deleted{s} using the true value of M. \added{For the point and interval estimation comparison, we include iterations in which these methods and those presented in the main text all produce finite estimates; as such the values for the original methods may differ slightly. Table \ref{mhat.eda} gives the median values of $\hat{M}$ for each DGP.}

\begin{figure}[t]
    \centering
    \includegraphics[width=\textwidth]{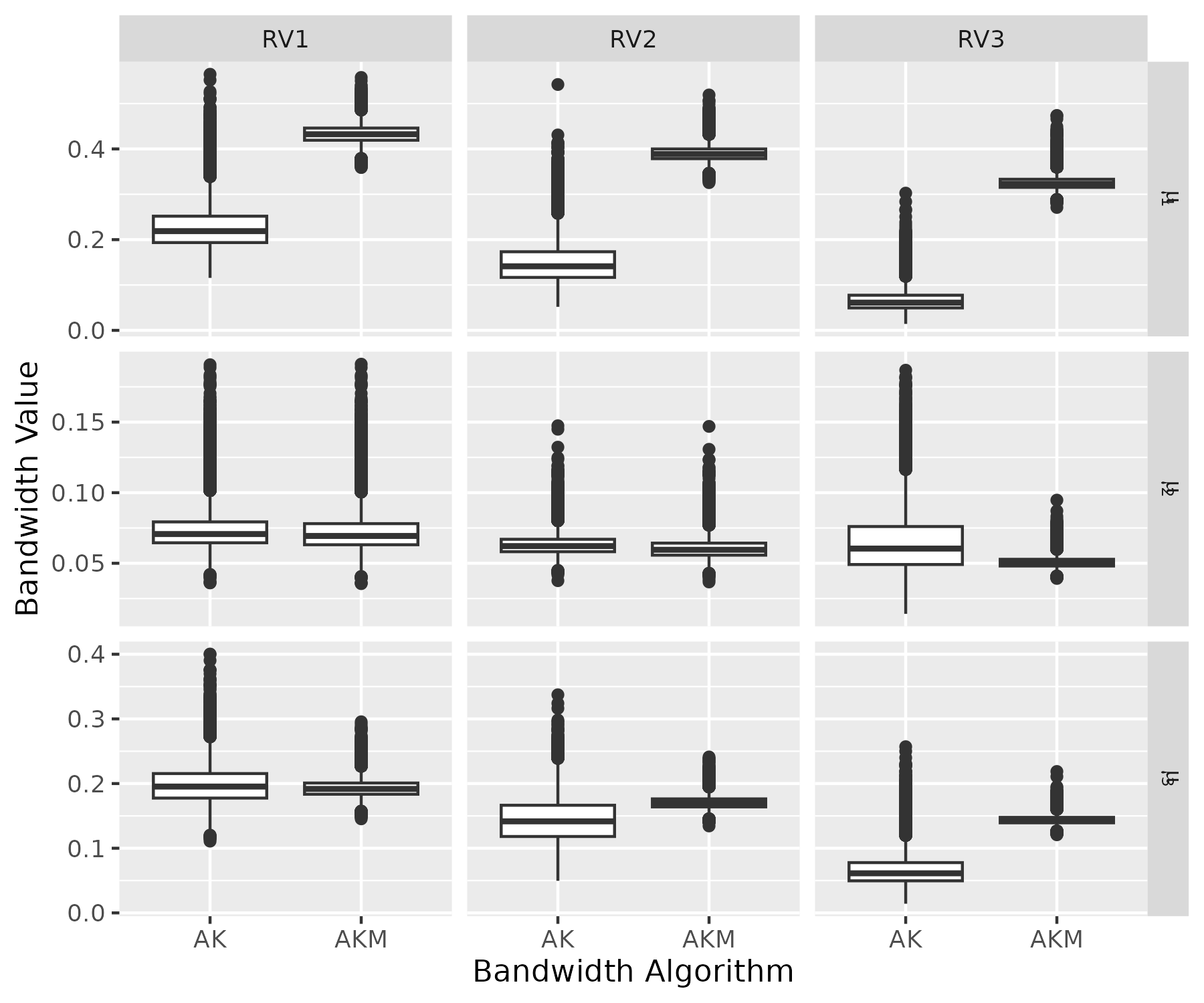}
    \caption{Bandwidth distributions for AK and AKM when $\bar{m}=27$.}
    \label{bw.sens}
\end{figure}

\deleted{There is naturally less variability in bandwidth values when using the consistent value of $M$ rather than estimating $\hat{M}$ each time, which we see in Figure \ref{bw.sens} for $\bar{m}=20$.} The relative size of the bandwidths\added{, whose distributions for $\bar{m}=27$ are given in Figure \ref{bw.sens},} depends on the relative values of $M$ and $\hat{M}$\deleted{, which was given in Table \ref*{mhat.eda} in the main text}. Larger values of $\hat{M}$ relative to $M$, as in $\mu_1$ and $\mu_3$, lead to smaller bandwidth values for AK\deleted{1} than AKM\deleted{1}. This is not surprising, because if we suspect larger curvature in the underlying mean function we would not want to include values as far away from the cutoff as we would if we suspected smaller curvature. We see the opposite trend in the $\mu_2$ setting, although the relative differences are smaller. \deleted{We see very similar results when considering AK2 and AKM2, and these trends continue}\added{Similar relationships exist} for \replaced{the}{other} values of $\bar{m}$ \added{not shown}. \added{There is naturally less variability in bandwidth values when using the consistent value of $M$ rather than estimating $\hat{M}$ each time.}

\begin{table}[t]
\caption{Median $\hat{M}$ values for different DGPs. True values of $M$ provided for comparison.}
\begin{tabulary}{\textwidth}{LRRRRRRRRRRRRRRRR}
\toprule
\multicolumn{1}{l}{} &  & \multicolumn{5}{c}{RV1} & \multicolumn{5}{c}{RV2} & \multicolumn{5}{c}{RV3} \\ 
\cmidrule(lr){3-7} \cmidrule(lr){8-12} \cmidrule(lr){13-17}
\multicolumn{1}{l}{} & $M$ & 10 & 21 & 27 & 44 & 57 & 10 & 21 & 27 & 44 & 57 & 10 & 21 & 27 & 44 & 57 \\ 
\midrule
$\mu_1$ & $6$ & $23$ & $14$ & $12$ & $9$ & $8$ & $90$ & $33$ & $26$ & $18$ & $15$ & $763$ & $200$ & $144$ & $86$ & $68$ \\ 
$\mu_2$ & $233$ & $211$ & $220$ & $222$ & $224$ & $225$ & $203$ & $207$ & $210$ & $214$ & $216$ & $762$ & $203$ & $148$ & $94$ & $78$ \\ 
$\mu_3$ & $16$ & $25$ & $17$ & $16$ & $14$ & $13$ & $91$ & $33$ & $26$ & $19$ & $16$ & $762$ & $200$ & $144$ & $86$ & $68$ \\ 
\bottomrule
\end{tabulary}
\label{mhat.eda}
\end{table}

The AKM\deleted{1 and AKM2} bandwidth algorithm\replaced{ has}{s have} near universal success rates similar to those of AK\deleted{1 and AK2}. The interval success rates are once again tied to the bandwidth size. Table \ref{tabsens} shows the interval success rates for AKM\deleted{1} and AK\deleted{M2} for \replaced{all 9 DGPs}{RV2}. For $\mu_1$ and $\mu_3$, the larger bandwidths of AKM\deleted{1} \deleted{and AKM2 led}\added{lead} to higher success rates than those of AK\deleted{1} \deleted{and AK2 (see Table \ref{tab3})}\added{for all study sizes}. For $\mu_2$, the methods using the true $M$ \replaced{have}{had} lower success rates\added{ except for RV3}. \deleted{This trend holds for the other running variable distributions as well.} Once a bandwidth has been calculated, using FLCIM rather than FLCI does not have a meaningful effect on the interval success rates.

\begin{table}[t]
\caption{Interval estimate success rates (in percents) for RV2 for bandwidth algorithms using the true value of M.}
\begin{tabularx}{\textwidth}{l|IIIIIIIIII}
\toprule
\multicolumn{1}{l}{} &  & \multicolumn{3}{c}{$\bar{m}=10$} & \multicolumn{3}{c}{$\bar{m}=21$} & \multicolumn{3}{c}{$\bar{m}=27$} \\ 
\cmidrule(lr){3-5} \cmidrule(lr){6-8} \cmidrule(lr){9-11}
\multicolumn{1}{l}{} & BW & RBC & CV & FLCI & RBC & CV & FLCI & RBC & CV & FLCI \\ 
\midrule
\multicolumn{1}{l}{$\mu_1$} \\ 
\midrule
RV1 & AK & $89.93$ & $90.06$ & $100.00$ & $99.96$ & $99.96$ & $100.00$ & $100.00$ & $100.00$ & $100.00$ \\ 
 & AKM & $100.00$ & $100.00$ & $100.00$ & $100.00$ & $100.00$ & $100.00$ & $100.00$ & $100.00$ & $100.00$ \\ 
RV2 & AK & $80.47$ & $80.87$ & $98.71$ & $99.85$ & $99.85$ & $100.00$ & $99.99$ & $99.99$ & $100.00$ \\ 
 & AKM & $99.61$ & $99.61$ & $98.71$ & $100.00$ & $100.00$ & $100.00$ & $100.00$ & $100.00$ & $100.00$ \\ 
RV3 & AK & $64.67$ & $65.28$ & $91.03$ & $99.25$ & $99.25$ & $100.00$ & $99.94$ & $99.94$ & $100.00$ \\ 
 & AKM & $96.26$ & $96.26$ & $91.03$ & $100.00$ & $100.00$ & $100.00$ & $100.00$ & $100.00$ & $100.00$ \\ 
\midrule
\multicolumn{1}{l}{$\mu_2$} \\ 
\midrule
RV1 & AK & $52.51$ & $53.62$ & $99.39$ & $84.68$ & $84.84$ & $100.00$ & $93.28$ & $93.33$ & $100.00$ \\ 
 & AKM & $51.01$ & $52.12$ & $99.39$ & $83.02$ & $83.14$ & $100.00$ & $92.91$ & $92.96$ & $100.00$ \\ 
RV2 & AK & $66.41$ & $67.11$ & $98.57$ & $97.07$ & $97.09$ & $100.00$ & $99.35$ & $99.36$ & $100.00$ \\ 
 & AKM & $66.03$ & $66.67$ & $98.57$ & $96.26$ & $96.27$ & $100.00$ & $99.17$ & $99.17$ & $100.00$ \\ 
RV3 & AK & $64.33$ & $64.91$ & $91.03$ & $99.27$ & $99.28$ & $100.00$ & $99.96$ & $99.96$ & $100.00$ \\ 
 & AKM & $85.88$ & $86.01$ & $91.03$ & $99.81$ & $99.81$ & $100.00$ & $99.99$ & $99.99$ & $100.00$ \\ 
\midrule
\multicolumn{1}{l}{$\mu_3$} \\ 
\midrule
RV1 & AK & $88.55$ & $88.70$ & $100.00$ & $99.94$ & $99.94$ & $100.00$ & $100.00$ & $100.00$ & $100.00$ \\ 
 & AKM & $94.59$ & $94.62$ & $100.00$ & $99.97$ & $99.97$ & $100.00$ & $100.00$ & $100.00$ & $100.00$ \\ 
RV2 & AK & $80.23$ & $80.66$ & $98.71$ & $99.86$ & $99.86$ & $100.00$ & $100.00$ & $100.00$ & $100.00$ \\ 
 & AKM & $98.05$ & $98.06$ & $98.71$ & $100.00$ & $100.00$ & $100.00$ & $100.00$ & $100.00$ & $100.00$ \\ 
RV3 & AK & $64.64$ & $65.25$ & $91.03$ & $99.30$ & $99.31$ & $100.00$ & $99.96$ & $99.96$ & $100.00$ \\ 
 & AKM & $95.93$ & $95.93$ & $91.03$ & $100.00$ & $100.00$ & $100.00$ & $100.00$ & $100.00$ & $100.00$ \\ 
\bottomrule
\end{tabularx}
\label{tabsens}
\end{table}

\begin{figure}[t!]
    \centering
    \includegraphics[width=\textwidth]{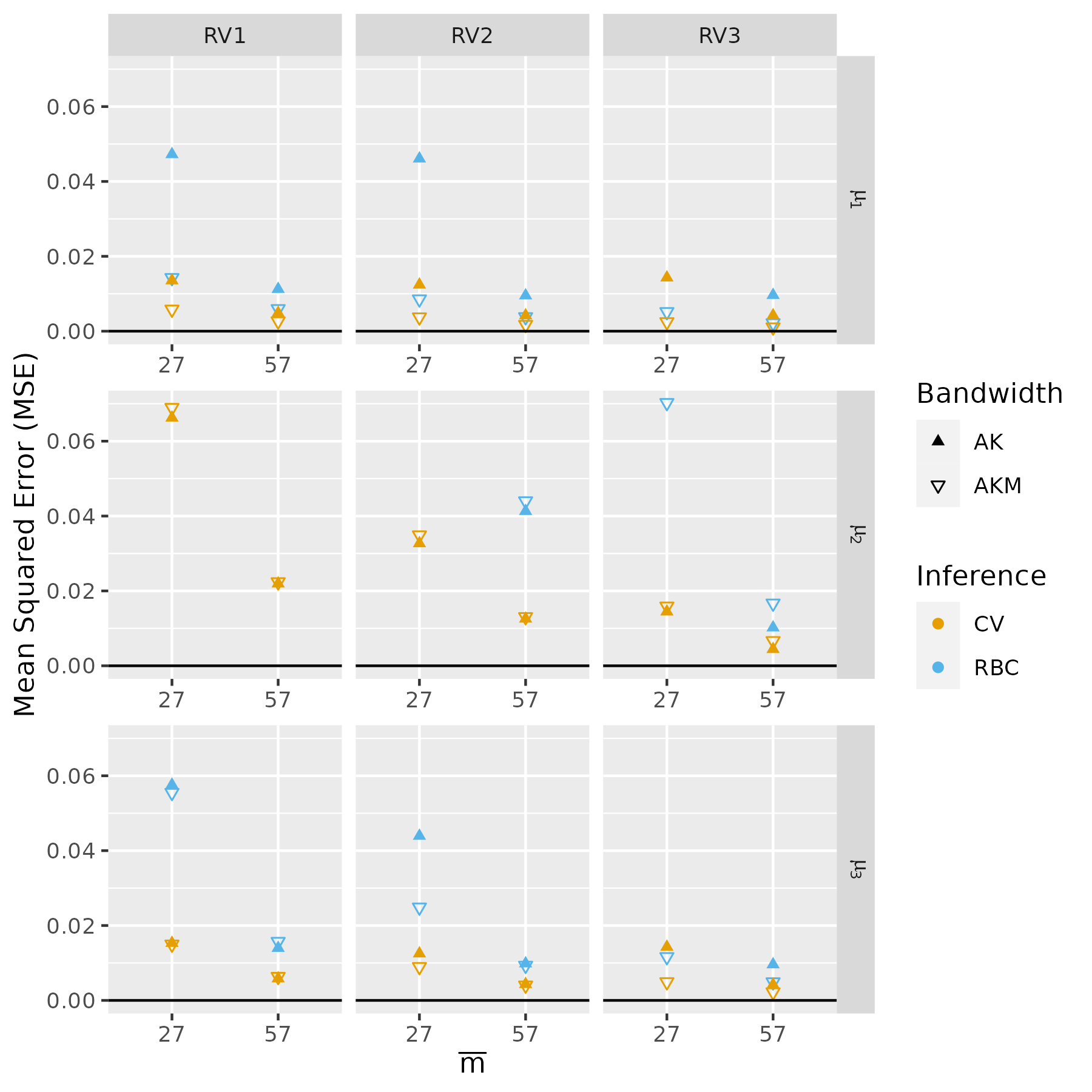}
    \caption{Mean squared error values for methods using AK and AKM algorithms. Values based on iterations in which estimates were finite for all methods, including those not shown but considered in the main text. This includes at least 99$\%$ of all iterations except for RV1$\mu_2$ for $\bar{m}=27$ (90$\%$). The RV1$\mu_2$ graph omits values when $\bar{m}=27$ for AK/RBC (393) and AKM/RBC (1177), and when $\bar{m}=57$ for AK/RBC (0.19) and AKM/RBC (0.23). The RV2$\mu_2$ graph omits values when $\bar{m}=27$ for AK/RBC (204) and AKM/RBC (204). The RV3 graph omits values when $\bar{m}=27$ for AK/RBC for all $\mu$ (0.21). The maximum Monte Carlo standard errors (MCSE) for the estimates shown is 0.004.}
    \label{scattersens}
\end{figure}

For methods using CV inference, there \replaced{are some}{is a relatively} small difference\added{s} in MSE values for methods using AK\deleted{1} and those using AKM\deleted{1} \added{for several DGPs}, as seen in Figure \ref{scattersens}\added{, particularly for the smaller study size of $\bar{m}=27$}. \added{Naturally the DGPs with larger MSE differences tend to be the ones in which the median value of $\hat{M}$ differs substantially from the true value of M. Note that the difference is not always in the same direction. For $\mu_1$ and $\mu_3$, MSE values tend to be smaller for AKM, but the opposite is true for $\mu_2$. Recall that for these sample sizes the algorithm tends to underestimate the quite large true value of $M$ for $\mu_2$. However, this maximum curvature occurs at the edge of the support of the running variable, and a small data set may not contain values that extreme. Thus it is possible that an estimation based on the data may be closer to the curvature of the running variable in the actual range of the data than the true value based on the entire support, and this may be contributing to the slightly better performance of the AK method here.}

\replaced{The differences in MSE values tend to be larger but in the same direction when considering RBC instead of CV inference. For the larger study size of $\bar{m}=57$ the MSE values tend to be more similar. For larger study sizes the algorithm will naturally do a better job of estimating the true value of $M$ and we would expect to see less of a benefit of using $M$ \textit{a priori}.}{However the change is larger for methods using RBC inference, especially for smaller sample sizes. In those cases the methods using the true value of $M$ fair better, as would be expected. Even with this improvement the CV methods tend to outperform the RBC methods in terms of point estimation. A comparison of methods using AK2 and AKM2 yields similar results.}

\begin{figure}[t!]
    \centering
    \includegraphics[width=\textwidth]{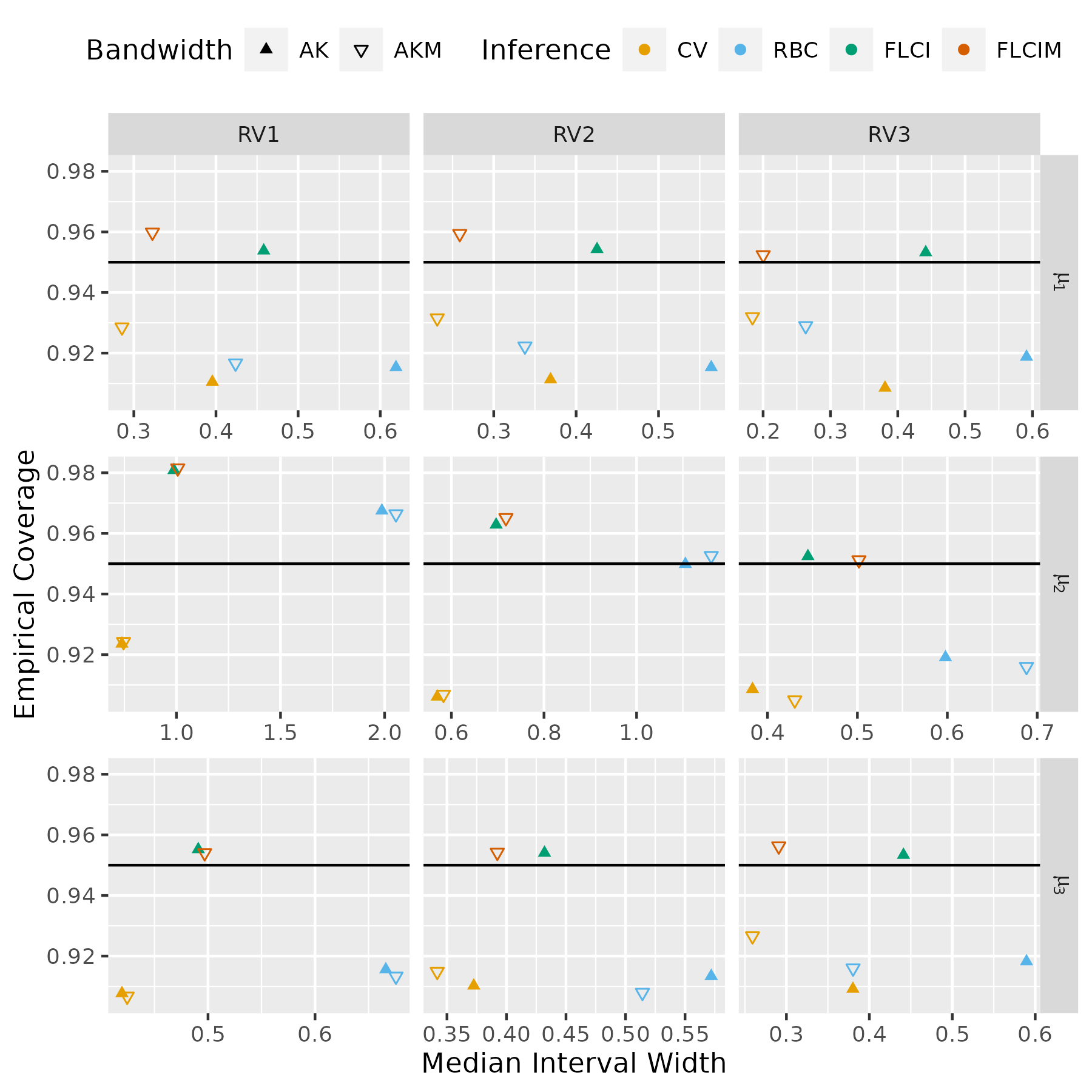}
    \caption{Empirical coverage and median interval widths for select methods using AK and AKM algorithms when $\bar{m}=27$. Values based on iterations in which estimates were finite for all methods, including those not shown but considered in the main text. This includes at least 99$\%$ of all iterations except for RV1$\mu_2$ for $\bar{m}=27$ (90$\%$). The maximum MCSE value for these estimates is 0.001.}
    \label{cvsens}
\end{figure}

The effect of using the true value of $M$ on interval estimation \added{again} varies \deleted{considerably} depending on DGP. Figure \ref{cvsens} shows the median interval widths and coverage for all DGPs and methods using AK1 and AKM1 when $\bar{m}=$\replaced{27}{20}. It includes all four types of inference. However, it excludes AK\deleted{1}/FLCIM and AKM\deleted{1}/FLCI because in practice the same derivative bound would be used for both bandwidth and interval calculations. For $\mu_2$, there is little difference between AK\deleted{1}/FLCI and AKM\deleted{1}/FLCIM\added{ except for RV3, where the former has narrower median widths and slightly better coverage despite using an estimated value of M}. However for $\mu_1$ and $\mu_3$, using the true value of $M$ \added{tends to} lead\deleted{s} to narrower intervals for about the same coverage. \replaced{Similar relationships exist between AK/CV and AKM/CV, although the differences in coverage are often more pronounced.}{For methods using CV inference, using the true value of $M$ leads to better coverage and narrower intervals for $\mu_1$, but less of a difference for the rest. For methods using RBC inference, using the true value of $M$ leads to less coverage but narrower widths for $\mu_1$ and $\mu_3$.} For larger study sizes the trends are not meaningfully different.

Based on our sensitivity analysis using the true value of $M$ can provide a substantial benefit in certain situations, including some DGPs where the $\hat{M}$ estimates are not very close to the truth. \added{However in other situations it would appear that using a true, but very large, value of $M$ may lead to slightly worse performance}. If a researcher has strong evidence for a derivative bound, they can certainly use that value. However, it would seem to be a good idea to also use the data driven $\hat{M}$ value in practice. If the two derivative bounds do not match, further investigation into the data should probably take place.

\printbibliography
\end{refsection}

\end{document}